\documentclass[twocolumn,english,aps,pra,10pt,superscriptaddress,floatfix]{revtex4-2}

\usepackage{times}
\usepackage{bm}
\usepackage{graphicx}
\usepackage{amssymb,amsfonts,amsmath,amsbsy,bm,t1enc,latexsym}
\usepackage{float}
\usepackage[colorlinks=true,citecolor=blue,linkcolor=magenta]{hyperref}
\usepackage[markup=nocolor, authormarkupposition=left]{changes}
\usepackage[english]{babel}
\usepackage{url}%

\usepackage{color,soul}
\usepackage{textcomp}
\usepackage[]{lipsum}

\newcommand{\SiN}[0]{Si$_3$N$_4$}

\begin{document}
	
	\title{Soliton Microcomb Generation in a III-V Photonic Crystal Cavity}
	 
\author{Alberto Nardi}
\thanks{These authors contributed equally to this work.}
\affiliation{IBM Research Europe, Zurich, S\"{a}umerstrasse 4, R\"{u}schlikon, CH-8803 Switzerland}
\affiliation{Institute of Physics, Swiss Federal Institute of Technology Lausanne (EPFL), CH-1015 Lausanne, Switzerland}
	
\author{Alisa Davydova}
\thanks{These authors contributed equally to this work.}
\affiliation{Institute of Physics, Swiss Federal Institute of Technology Lausanne (EPFL), CH-1015 Lausanne, Switzerland}
\affiliation{Center of Quantum Science and Engineering (EPFL), CH-1015 Lausanne, Switzerland}

\author{Nikolai Kuznetsov}
\thanks{These authors contributed equally to this work.}
\affiliation{Institute of Physics, Swiss Federal Institute of Technology Lausanne (EPFL), CH-1015 Lausanne, Switzerland}
\affiliation{Center of Quantum Science and Engineering (EPFL), CH-1015 Lausanne, Switzerland}

\author{Miles H. Anderson}
\affiliation{Institute of Physics, Swiss Federal Institute of Technology Lausanne (EPFL), CH-1015 Lausanne, Switzerland}
\affiliation{Center of Quantum Science and Engineering (EPFL), CH-1015 Lausanne, Switzerland}

\author{Charles M\"{o}hl}
\affiliation{IBM Research Europe, Zurich, S\"{a}umerstrasse 4, R\"{u}schlikon, CH-8803 Switzerland}

\author{Johann Riemensberger}
\affiliation{Institute of Physics, Swiss Federal Institute of Technology Lausanne (EPFL), CH-1015 Lausanne, Switzerland}
\affiliation{Center of Quantum Science and Engineering (EPFL), CH-1015 Lausanne, Switzerland}

\author{Paul Seidler}
\email[]{pfs@zurich.ibm.com}
\affiliation{IBM Research Europe, Zurich, S\"{a}umerstrasse 4, R\"{u}schlikon, CH-8803 Switzerland}

\author{Tobias J. Kippenberg}
\email[]{tobias.kippenberg@epfl.ch}
\affiliation{Institute of Physics, Swiss Federal Institute of Technology Lausanne (EPFL), CH-1015 Lausanne, Switzerland}
\affiliation{Center of Quantum Science and Engineering (EPFL), CH-1015 Lausanne, Switzerland}

	\maketitle
	
	\subsection*{Abstract}
	\textbf{Photonic crystals, material structures in which the dielectric function varies periodically in one, two, or three dimensions, can provide exquisite control over the propagation and confinement of light~\cite{de_la_rue_photonic_2012, joannopoulos_photonic_2008, lourtioz_photonic_2005}.  By tailoring their band structure, exceptional optical effects can be achieved, such as slow light propagation or, through the creation of photonic bandgaps, optical cavities with both a high quality factor and a small mode volume~\cite{akahane_high-q_2003}. 	
	Photonic crystal cavities have been used to realize compact nano-lasers and achieve strong coupling to quantum emitters, such as semiconductor quantum dots~\cite{schererPhotonicCrystalsConfining2002,reithmaier_strong_2004}, color centers~\cite{evans_photon-mediated_2018}, or  cold atoms~\cite{douglasQuantumManybodyModels2015, gobanAtomLightInteractions2014}. 
	A particularly useful attribute of photonic crystals is the ability to create chirped mirrors.
	Chirping has underpinned advances in ultra-fast lasers based on bulk mirrors~\cite{kartner_design_1997,xu_design_2016, jungSelfstarting65fsPulses1997}, but has yet to be fully exploited in integrated photonics, where it could 
	provide a means to engineer otherwise unattainable dispersion profiles for a range of nonlinear optical applications, including soliton frequency comb generation. 
	The vast majority of integrated resonators for frequency combs make use of microring geometries, where only waveguide width and height are varied to engineer dispersion.	
	Recently, generation of frequency combs has been demonstrated with one-dimensional photonic crystal cavities made of silicon nitride~\cite{yu_photonic-crystal-reflector_2019,wildi_soliton_2022}, but the low index contrast prevents formation of broad soliton combs.	
	Here we overcome these challenges by using a photonic-crystal Fabry-P\'{e}rot resonator made of gallium phosphide, a material exhibiting negligible two-photon absorption at telecommunication wavelengths, a high refractive index, and a Kerr nonlinearity 200 times larger than that of silicon nitride.  We employ chirped photonic crystal mirrors to provide anomalous dispersion. 
	With subharmonic pulsed pumping at an average power of 23.6~mW, we are able to access stable dissipative Kerr frequency combs.  We demonstrate soliton formation with a 3-dB bandwidth of 3.0~THz, corresponding to a pulse duration of 60~fs. 
	This approach to cavity design offers nearly arbitrary dispersion engineering over the optical transparency window of the nonlinear material.}
	
	\subsection*{Introduction}\label{intro}
	A temporal dissipative Kerr soliton (DKS)~\cite{leo_temporal_2010,herr_temporal_2014,kippenberg_dissipative_2018,gaeta_photonic-chip-based_2019} represents a stationary attractor of coherently driven optical cavities possessing a third order nonlinearity. 
	The system is described by the Lugiato-Lefever equation (LLE)~\cite{lugiato_spatial_1987}, and a stable bright soliton emerges when anomalous group velocity dispersion (GVD) compensates the optical nonlinearity and the Kerr parametric gain balances the resonator loss. Today it is well understood that driven dissipative nonlinear resonators support the formation of a range of coherent dissipative structures that includes bright solitons in the anomalous GVD regime, but also switching waves forming dark solitons (also referred to as platicons) in the normal GVD regime~\cite{kivshar_dark_1998,godey_stability_2014,liang_generation_2014,xue_mode-locked_2015,huang_mode-locked_2015}, as well as zero-dispersion solitons~\cite{anderson_zero_2022}. 
	Such structures can be generated using a continuous-wave (c.w.)~\cite{herr_temporal_2014} laser or with a pulsed optical drive ~\cite{obrzud_temporal_2017}, which increases the nonlinear conversion efficiency while reducing average power requirements. Bright dissipative Kerr solitons, in particular, have been utilized in numerous system-level demonstrations, including massively parallel coherent communications~\cite{marin-palomo_microresonator-based_2017,fulop_high-order_2018}, coherent frequency-modulated continuous-wave (FMCW) light detection and ranging (LiDAR)~\cite{riemensberger_massively_2020}, dual-comb coherent LiDAR~\cite{lukashchukDualChirpedMicrocomb2022}, low noise microwave generation~\cite{lucas_ultralow-noise_2020,liuPhotonicMicrowaveGeneration2020}, dual-comb spectroscopy~\cite{suh_microresonator_2016}, astronomical spectrograph calibration~\cite{obrzud_microphotonic_2019,suh_searching_2019},  optical atomic clocks~\cite{newman_architecture_2019}, and neuromorphic computing~\cite{feldmannParallelConvolutionalProcessing2021}.	
	
	Among the numerous material platforms in which DKS~\cite{kippenberg_dissipative_2018} generation has been observed, silicon nitride (\SiN) microring resonators have emerged as one of the most widely used platforms, including for nearly all system-level applications listed above.  Si$_3$N$_4$ waveguides have low propagation loss (leading to large quality factors and, in turn, low thresholds for soliton initiation), the ability to handle high power, weak thermal effects, weak competing Raman or Brillouin scattering, and foundry-level maturity.
	Although octave-spanning combs have been demonstrated using dual dispersive wave generation in such microring resonators, they do not offer many degrees of freedom for dispersion engineering; typically, waveguide thickness and width are the only available parameters.
	Cavities with chirped mirrors provide an attractive alternative, as first demonstrated by Szipöcs et al. in 1994~\cite{szipocs_chirped_1994} with non-periodic coatings of bulk mirrors, an approach that remains one of the preeminent dispersion control methods in ultrafast laser technology~\cite{kartner_design_1997, xu_design_2016, jungSelfstarting65fsPulses1997}. 
	
	Dispersion engineering with chirped mirrors can be realized in on-chip integrated devices
	with photonic crystal Fabry-P\'{e}rot (PC-FP) resonators.  Photonic crystal structures have recently gained attention for both Fabry-P\'{e}rot cavities~\cite{yu_photonic-crystal-reflector_2019, wildi_soliton_2022} and ring resonators~\cite{lucas_tailoring_2022,yuContinuumBrightDarkpulse2022} because of the versatility provided by their multiple geometric degrees of freedom.  
	A photonic crystal reflector (PCR) can be designed to target a photonic bandgap centered around a specific frequency, and, using the chirped-mirror concept, the cavity length can be made frequency dependent to tailor the dispersion profile. 
	Dispersion engineering with photonic crystal structures has been implemented in Si$_3$N$_4$ PC-FP resonators and Kerr frequency combs generated in c.w. operation~\cite{yu_photonic-crystal-reflector_2019}. The low index contrast between Si$_3$N$_4$ and air (or silicon oxide, in the case of cladded devices) limits however the performance of the PCRs because the penetration depth is large, constraining the attainable dispersion profile.
	More recently, Wildi et al.~\cite{wildi_soliton_2022} have generated a narrowband DKS with a duration of 300~fs in a Si$_3$N$_4$ PC-FP resonator using a dual-pump scheme to counteract the thermo-refractive effect~\cite{zhang_sub-milliwatt-level_2019,r_niu_perfect_2021}. 
	In this case, non-chirped PCRs were employed with $\sim$100 unit cells and a reflection bandwidth of about 6~THz. 
	
	A more promising material for realizing high-reflectivity, chirped PCRs for the generation of broad DKSs is gallium phosphide (GaP). GaP possesses a large refractive index ($n_0>3$), a substantial $\chi^{(3)}$ nonlinearity, a non-zero $\chi^{(2)}$ nonlinearity due to its non-centrosymmetric crystal structure, and negligible two-photon absorption for wavelengths above $1.1$ $\mu$m.  Process technology for fabrication of  integrated GaP photonic devices is also now available~\cite{wilson_integrated_2020,schneider_gallium_2018,honl_highly_2018}. GaP has been employed in both one-dimensional~\cite{schneider_optomechanics_2019,honl_microwave--optical_2022} and two-dimensional~\cite{rivoire_gallium_2008,rivoire_second_2009,gan_high-resolution_2012} photonic crystals. Frequency combs  have been generated in ring resonators made of GaP~\cite{wilson_integrated_2020}, but DKS generation has remained elusive. 
	A strong thermo-refractive resonance shift for III-V materials often prevents DKS generation at room temperature with c.w. pumping~\cite{moille_dissipative_2020}. 
		
	Here we overcome this challenge and demonstrate DKS generation in a chirped-mirror photonic crystal cavity made of GaP using pulsed optical driving.
	The resonator is designed to ensure both high finesse
	($\mathcal{O}$($10^2 $)) 
	and anomalous group velocity dispersion.  We show that the concept of dispersion engineering with non-periodic photonic-crystal mirrors in high-index III-V materials
	expands the range of dispersion designs compared to what can be achieved in the ring resonator geometry, without compromising the optical quality factor.
	Indeed, we achieve record-high intrinsic optical quality factors (as high as $Q_0=$ 1.2$\times$10$^6$) and low propagation losses (0.85 dB~cm$^{-1}$) for the GaP material system, outperforming even the previously reported values for ring resonators of $Q=$~5.0$\times$10$^5$ and 1.2~dB~cm$^{-1}$, respectively~\cite{wilson_integrated_2020}.
	
		\begin{figure*}[]
		\centering
		\includegraphics[width=1\linewidth]{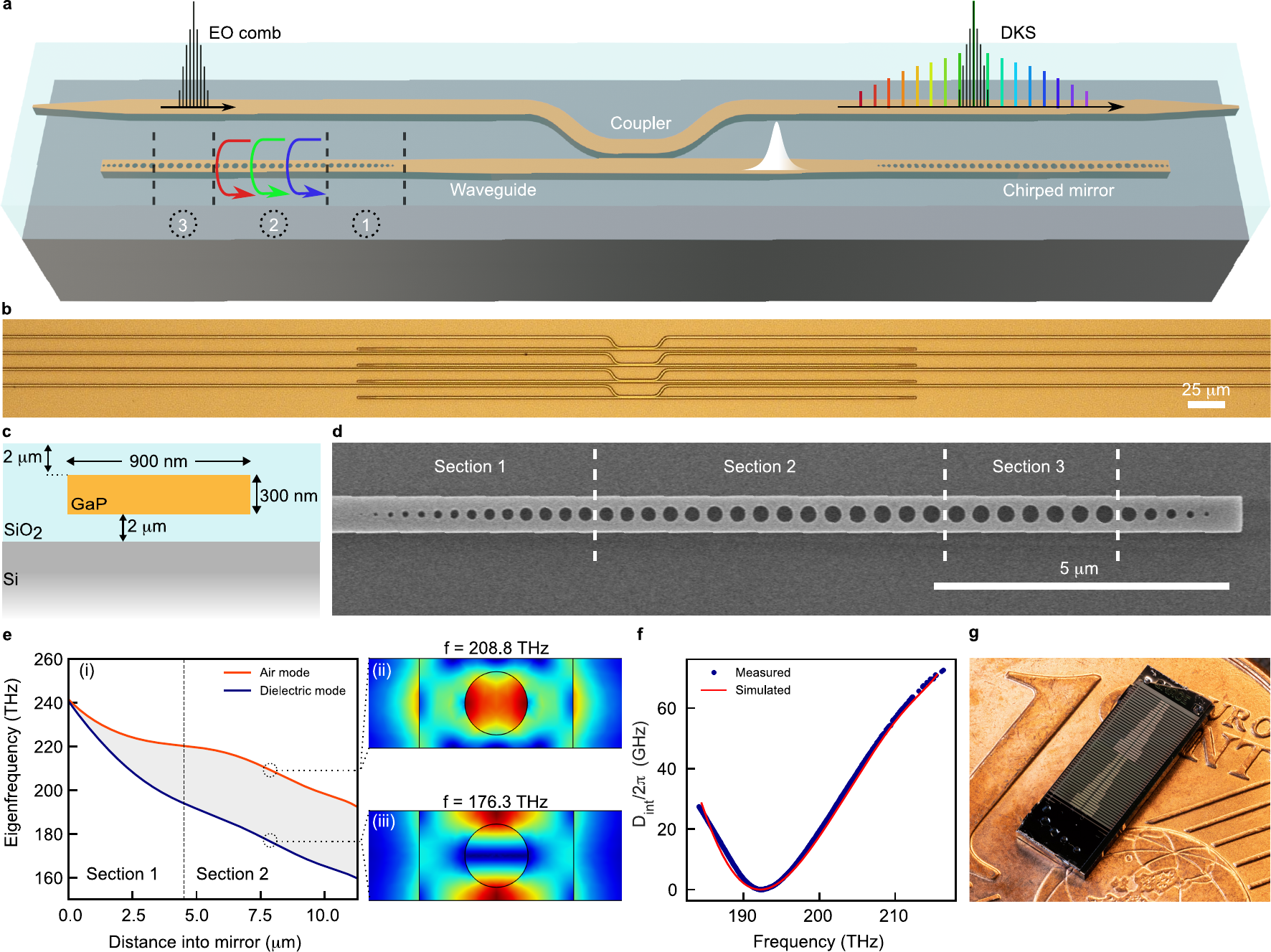}
		\caption{\textbf{Design of photonic-crystal Fabry-P\'{e}rot (PC-FP) cavity with chirped mirrors for dissipative Kerr soliton generation}.
			(a) Schematic of a PC-FP resonator made of GaP.
			An EO comb is coupled into the chip at the edge via inverse-tapered waveguides and evanescently into the resonator via a directional coupler. The resonator is composed of a straight waveguide with normal dispersion and PCRs at each end that contribute anomalous dispersion. Section \textit{1} of the PCR provides an adiabatic transition from the straight waveguide. In section \textit{2}, lower frequencies are reflected deeper into the PCR, resulting in anomalous dispersion. Section \textit{3} is composed of holes with fixed dimensions to enhance reflectivity at low frequencies. 
			(b) Optical microscope image of a group of devices showing dense integration.
			(c) Schematic of the GaP waveguide cross-section with dimensions (not drawn to scale). 
			(d) Scanning electron microscopy image of a fabricated PCR. 
			(e) Evolution of the transverse electric (TE) bandgap as a function of depth into the PCR. 
			(i) The bandgap is given by the dielectric mode (blue line) and air mode (orange line) at the edge of the first Brillouin zone.
			The bandgap is opened in the tapered section \textit{1}. In section \textit{2}, the bandgap is maintained at about 32~THz while its center frequency is chirped towards lower frequencies. 
			(ii, iii) Simulation of the norm of the electric field for TE air and dielectric modes, respectively, at the edge of the first Brillouin zone for a unit cell with bandgap center frequency of 192.6~THz (unit cell dimensions: thickness = 300~nm, width = 700~nm, hole pitch = 407~nm, hole radius = 143~nm).
			(f) Measured and simulated integrated dispersion of a device used to generate DKSs.
			(g) Photograph of fabricated sample on a 1 cent Euro coin.
		}\label{fig1}
	\end{figure*}

	\subsection*{Results}\label{sec2}
	
	\begin{figure*}[]%
		\centering
		\includegraphics[width=1\linewidth]{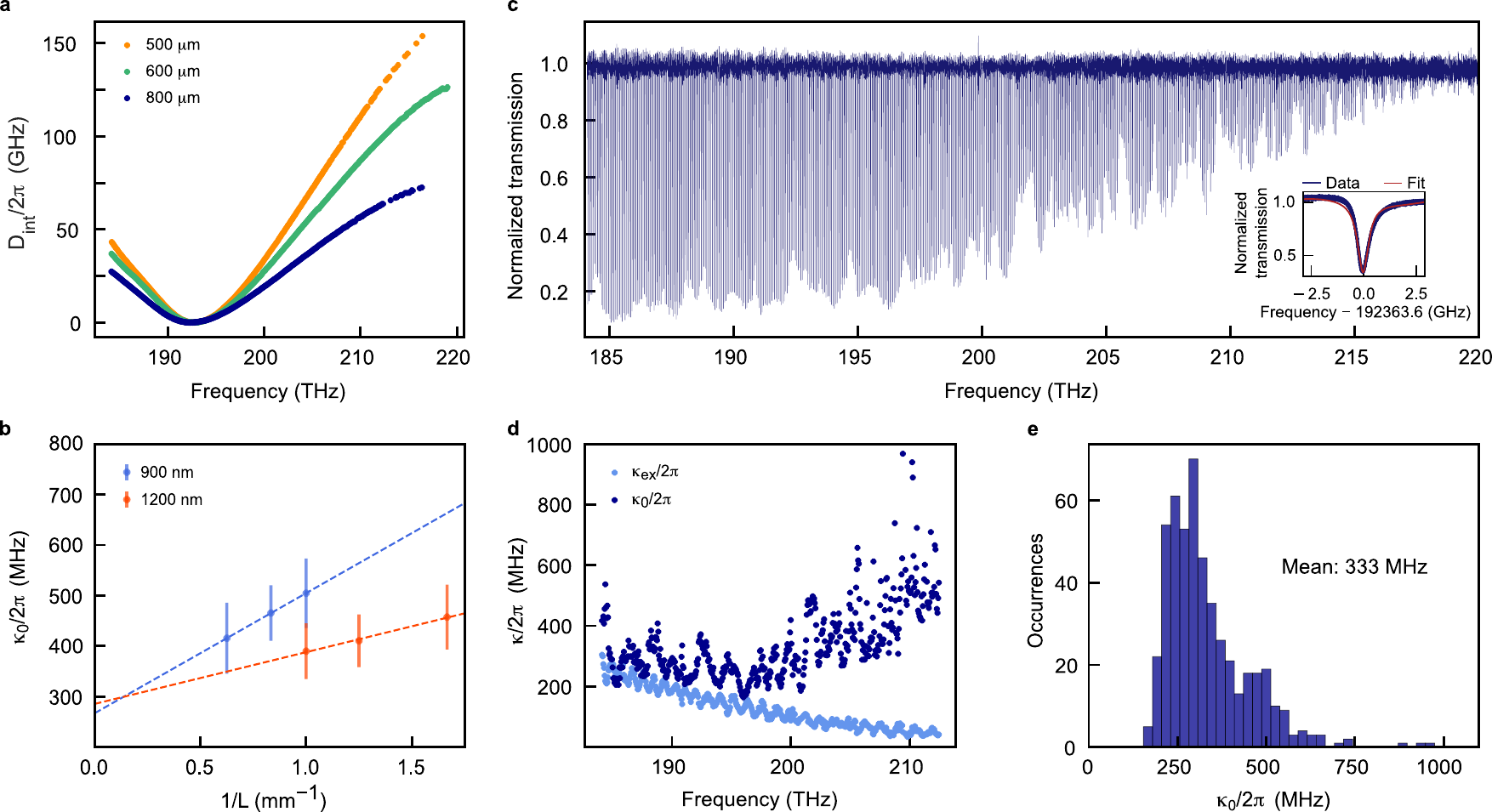}
		\caption{\textbf{Characterization of dispersion and losses.} 
			(a) Measured integrated dispersion of PC-FP resonators with various effective cavity lengths (500~$\mu$m, 600~$\mu$m and 800~$\mu$m) and width of 900~nm. 
			 The waveguide adds a normal dispersion contribution, hence the longer the waveguide the less anomalous the overall resonator dispersion. The blue curve corresponds to the device used for DKS generation.
			(b) Intrinsic loss rate versus the reciprocal of the effective cavity length for devices with waveguide widths of 900~nm and 1200~nm. The points are the average over all the resonances of several measured devices. Error bars are given by the measurement standard deviation. From the intercepts of the linear fits, we can calculate the waveguide propagation losses. 
			(c) Normalized transmission spectrum of the device used for DKS generation with effective cavity length of  800~$\mu$m and waveguide width of 900~nm. 
			Inset: Data and fit of the central resonance pumped with the EO comb, for which $\kappa_0/2\pi=366$~MHz, $\kappa_{\mathrm{ex}}/2\pi=132$~MHz, and $Q_0=5.26\times10^5$.
			(d) Intrinsic loss rate $\kappa_0/2\pi$ and external coupling rate $\kappa_{\mathrm{ex}}/2\pi$ for the device measured in (c). The external coupling rate decreases with frequency due to reduced evanescent coupling. The intrinsic loss rate oscillates because of the distributed photonic crystal mirrors. Outside the photonic bandgap, reduced reflectivity leads to increased intrinsic losses. On average, the values of $\kappa_0/2\pi$ correspond to intrinsic quality factors above $4\times 10^5$, reaching values as high as $Q_0 = 1.2\times 10^6$. 
			(e) Histogram of intrinsic loss values $\kappa_0/2\pi$ with a mean of 333~MHz.
		}\label{fig2}
	\end{figure*}

	\paragraph*{\textbf{Photonic crystal resonator design.}}
	The GaP resonator is composed of a straight waveguide cladded with a 2-$\mu$m-thick layer of SiO$_2$ with a PCR at each end, forming a monolithic Fabry-P\'{e}rot cavity
	(Fig.~\ref{fig1}a-b). Laser light is coupled in and out of the device through a bus waveguide with inverse tapers at the edges of the chip. Coupling to the resonator occurs evanescently via a directional coupler, the length of which is designed for critical coupling at the pump wavelength. The straight-waveguide section of the cavity has a cross-section of $300\times900$~nm$^2$ (Fig.~\ref{fig1}c) and thus has normal group velocity dispersion (see the Supplementary Information for more details on waveguide dispersion). The PCRs at the two ends consist of a series of holes in a waveguide with a cross-section of $300\times700$~nm$^2$, connected to the central straight waveguide section through a 40-$\mu$m-long taper (Fig.~\ref{fig1}). An initial section of 15 holes, in which the hole radius is linearly increased, gradually opens the photonic bandgap for the transverse electric mode to 32~THz and ensures that the transition between the central waveguide and the two PCRs is adiabatic. Importantly, the bandgap center frequency is then chirped down in frequency over 15 more periods so that lower frequencies are reflected deeper into the mirror (Fig.~\ref{fig1}e).
	In this way, anomalous dispersion and high reflectivity at all wavelengths of the chirped section are achieved. 
	(See the Supplementary Information for more details on PCR dispersion engineering).
	
	Fig.~\ref{fig1}f shows both  the measured and simulated integrated dispersion, 
	$D_{\mathrm{int}} = \omega_{\mu}-(\omega_0+D_1\mu) = \sum_{k=2}^{\infty}D_k\frac{\mu^k}{k!}$,	
	  of the PC-FP resonator used in this work to generate bright solitons. The experimentally measured free spectral range of the resonator is $\mathit{FSR}=D_1/2\pi=$~55.916~GHz at $\omega_0/2\pi=$~192.2~THz.
	The parameters $D_{k>1}$, describe higher orders of the dispersion.  The simulation considered dispersion parameters up to the 
	ninth order to account for oscillations in the group velocity dispersion profile typical of distributed chirped mirrors and is in good agreement with the experimental results (see \textit{Methods} for further details). Overall, the group velocity dispersion $\beta_2$ is anomalous around the pump frequency,  as $D_2 /2\pi = -c{D_{1}}^2 \beta_2/n_{\mathrm{g}} = 3.03$~MHz~>~0.  
	From the analysis of several devices, we find that the dispersion profile is quite reproducible and largely independent of the gap between the cavity and the bus waveguide in the directional coupler (see the Supplementary Information). 
	
		\begin{figure*}[]%
		\centering
		\includegraphics[width=1\linewidth]{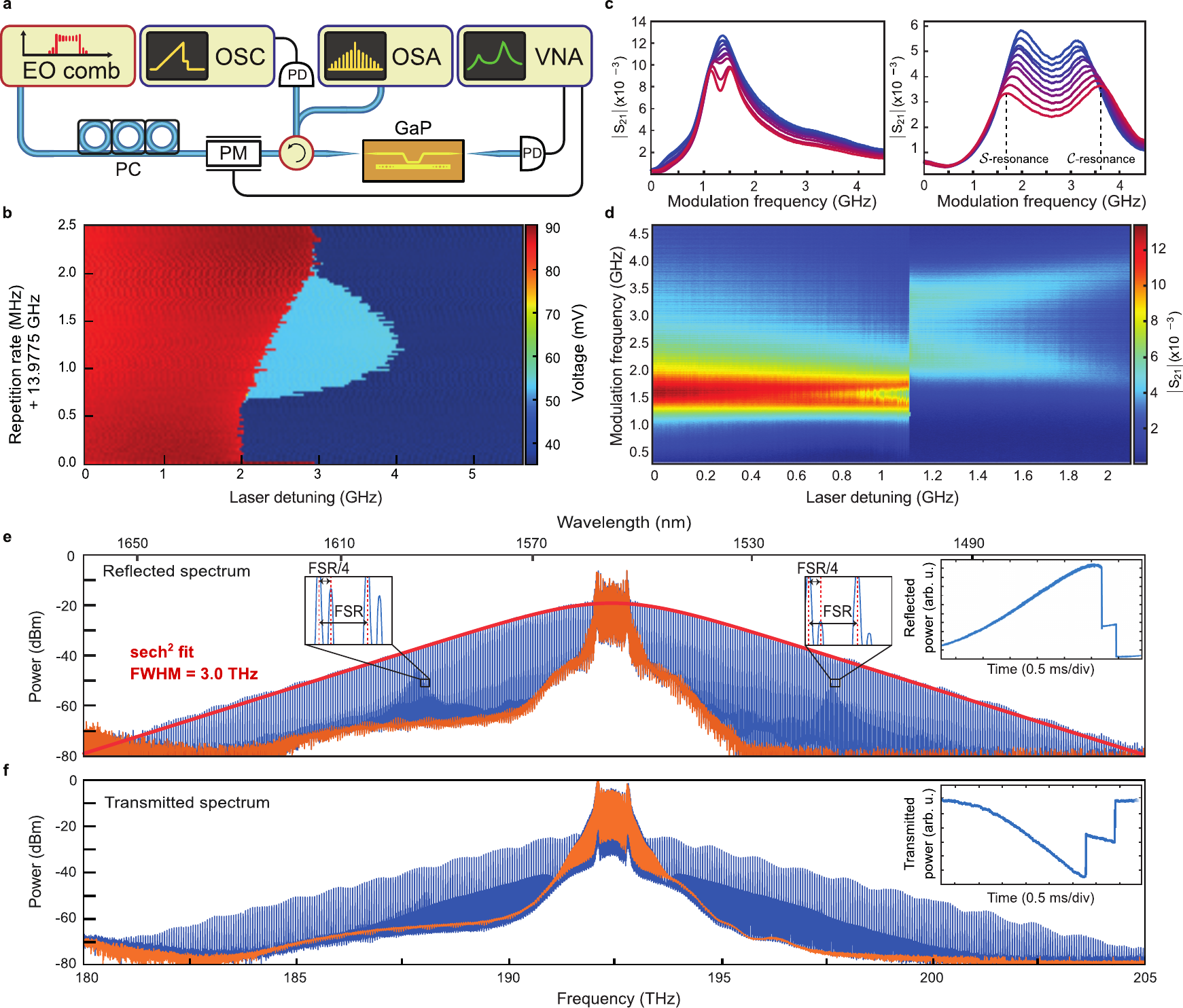}
		\caption{\textbf{Generation of dissipative Kerr solitons in a photonic-crystal Fabry-P\'{e}rot cavity with chirped mirrors} (a) Experimental apparatus for soliton generation.  A PC-FP resonator with FSR of 55.9~GHz is pumped with pulses from an EO comb.  The configuration shown is for recording spectra in reflection.  Alternatively, the optical spectrum analyzer (OSA) and oscilloscope (OSC) can be be connected to other side of the device to measure the transmitted light.  PC $-$ polarization controller; PM $-$ phase modulator; CIR $-$ optical circulator; PD $-$ photodiode; VNA $-$ vector network analyzer.
		(b) Compilation of oscilloscope traces measured in reflection as the laser is scanned through resonance for a range of repetition rates of the EO-comb source above threshold for DKS generation. The red region corresponds to the MI state. The light blue area is the soliton step. The dark blue shows the region beyond the range of soliton existence and off resonance.
    	(c) VNA response for the MI (left) and single-soliton (right) states. In the soliton state, the $\mathcal{S}$- and $ \mathcal{C} $-resonances move apart as the detuning increases, similar to the behavior of ring-type resonators. The plotted curves correspond to every tenth VNA trace in (d) as the laser is scanned across the resonance.
		(d) Evolution of the VNA response as a function of laser detuning. 
		(e) Spectra measured in reflection of the soliton state (blue) and off resonance (orange) for an average input power in the bus waveguide of $\sim$~70 mW. The two insets above the spectra show the positions of the additional weak lines offset from the main soliton comb lines (see the Supplementary Information for a detailed explanation). The inset on the right is an oscilloscope trace recorded as the laser is scanned through resonance.
	 	(f) Same as (e) measured in transmission.}
	 \label{fig3} 
	\end{figure*}

	\paragraph*{\textbf{Linear spectroscopy.}}
	We demonstrate the versatility of the PC-FP concept by examining a range of device geometries.
	By increasing the length of the straight waveguide section contributing normal dispersion while keeping the PCR design unchanged, the overall net dispersion becomes less anomalous (see Fig.~\ref{fig2}a).
	Increasing the width of the straight waveguide section from 900~nm to 1200~nm makes the overall dispersion only slightly less anomalous 
	($D_2/2\pi$ decreases by $\sim$10\%), but, more importantly, decreases the intrinsic loss rate $\kappa_0$; devices with a narrower waveguide have consistently higher loss.
	For a given waveguide width, the intrinsic loss rate can be approximated as	
	$\kappa_0 \approx \dfrac{c}{n_\mathrm{g}}\left(\dfrac{1-r}{L} +\alpha_\mathrm{wg}\right)$
	where $c$ is the speed of light, $n_{\mathrm{g}}$ is the effective group index,  $r$ is the reflection ratio of the PCRs and $\alpha_{wg}$ is the waveguide propagation loss (see Supplementary Information). We approximate the effective cavity length $L$ as the sum of the length of the straight waveguide section and the length of the two tapers, plus 20 $\mu$m to account for some penetration into the PCRs, from which we calculate an effective group index $n_{\mathrm{g}}=2\pi c/(2LD_1)=3.38$ for a resonator with an effective cavity length $L =800 \mu$m and width 900 nm ($n_{\mathrm{g}}=3.35$ for a width of 1200 nm). We can then extract the waveguide propagation loss $\alpha_{wg}$ from the intercept of a graph of $\kappa_0$ versus $1/L$ (Fig.~\ref{fig2}b).  The propagation loss is essentially the same ($\sim$0.85 dB cm$^{-1}$)
	within the error of the measurement for waveguides of width 900 nm and 1200 nm.  Given that the PCRs are of identical design, the difference in the slopes of the linear fits in Fig.~\ref{fig2}b suggests that the tapered transitions between the straight waveguide and the PCRs lead to different loss rates for the different widths.
	
	In Fig.~\ref{fig2}c, we report the normalized transmission spectrum 
	for the device used for DKS generation. The raw transmission spectrum is provided in the Supplementary Information. The spectrum is filtered with a 100-GHz-wide Savitsky-Golay filter to remove interference fringes due to reflections from the chip facets and to highlight the properties of the PCRs. We achieve a lensed-fiber-to-chip coupling loss of 4.8~dB per facet. The bandwidth of the PCRs is clearly evident in the extinction ratio of the resonances across the spectrum. As expected, starting around 200~THz, the PCRs exhibit lower reflectivity, \textit{i.e.}, the intrinsic loss of the cavity $\kappa_0$ is lower in the frequency range covered by the moving center frequency of the chirped mirrors (Fig.~\ref{fig2}d).
	In contrast, the extrinsic coupling loss $\kappa_{\mathrm{ex}}$ decreases monotonically with frequency, as expected for a directional coupler.  
	The distribution of $\kappa_0$ values (Fig.~\ref{fig2}e) is relatively narrow with a mean intrinsic loss rate of $\kappa_0/2\pi=333$~MHz for the 499 resonances measured.  
	The lowest value achieved is $\kappa_0/2\pi=$ 163.5 MHz at 196~THz, corresponding to an intrinsic quality factor of  $Q_0=1.2\times10^6$, the highest value reported to date for an integrated GaP microresonator. 
	It is worth noting that the value of $\kappa_0$ seems to gradually oscillate as a function of frequency in the high reflectivity region (Fig.~\ref{fig2}d), which we attribute to the discrete nature of the distributed PCRs. The smaller and faster oscillations in $\kappa_{\mathrm{ex}}$ are due to the relatively long length of the directional coupler (25~$\mu m$) required to couple sufficient light between the cavity and the bus waveguide.

	\begin{figure*}[]%
		\includegraphics[width=1\textwidth]{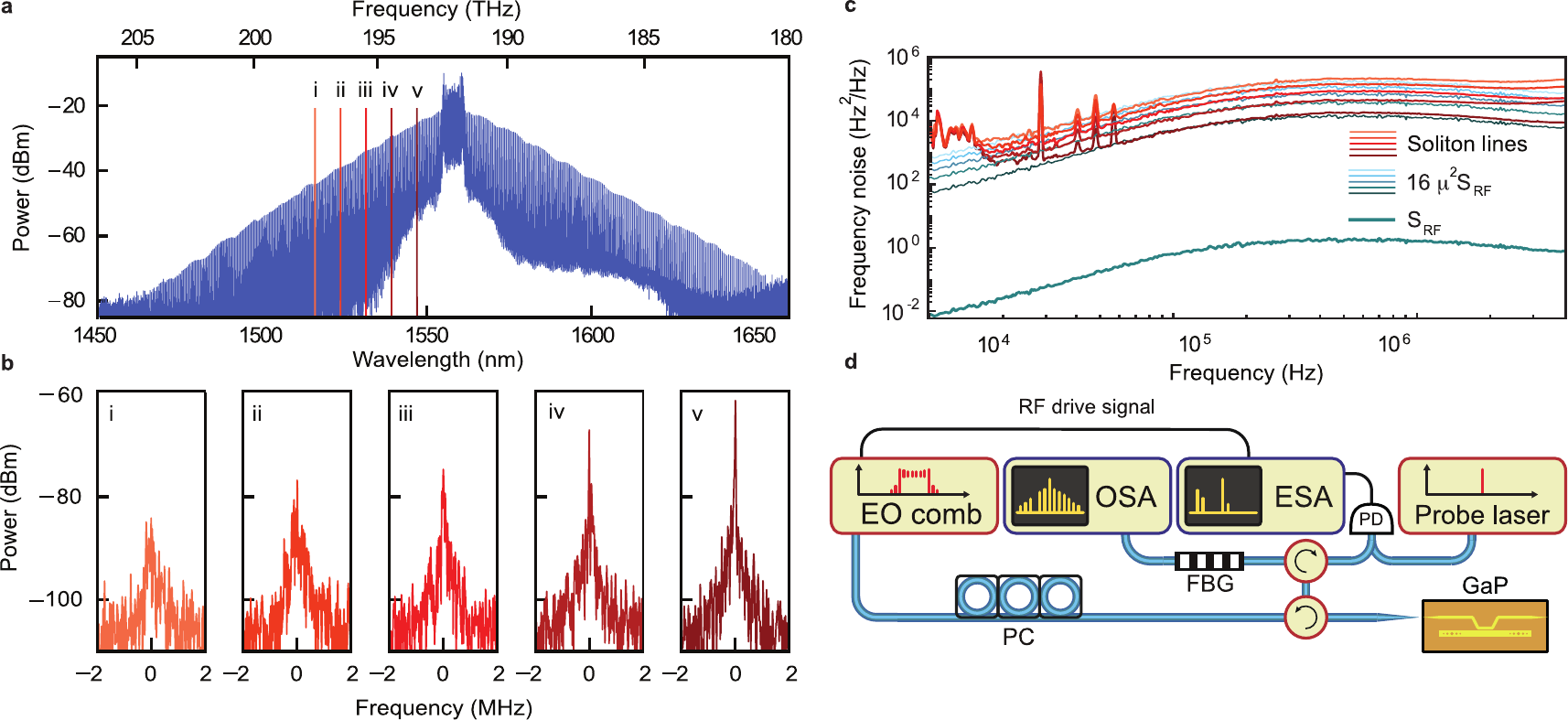}
		\caption{\textbf{Noise characteristics of the single-soliton state.} 
			(a) Soliton spectrum with the comb lines used in the heterodyne beatnote measurements indicated. 
			(b) Beatnotes of individual soliton lines with mode numbers (i) $\mu=77$, (ii) $\mu=63$, (iii) $\mu=49$, (iv) $\mu=36$, and (v) $\mu=22$. 
			(c) Frequency noise spectra of the RF source for the EO comb (teal) and the measured soliton lines (red gradient) with mode numbers $\mu$ as in (b). 
			The teal-gradient lines are given by $16\mu^2 S_{\mathrm{RF}}(f)$.
			The measured frequency noise of the corresponding soliton lines validates the noise multiplication relation $S_{\mathrm{\nu}}^{(\mu)}(f)=16\mu^2 S_{\mathrm{RF}}(f)$, where the factor of 16 is due to $f_\mathrm{rep}~$=~$\mathrm{FSR}/4$.
			(d) Experimental apparatus configured for noise measurements: 
			The fiber Bragg grating (FBG) reflects a comb line, which beats with the probe laser. A two-channel balanced photodiode (PD) collects the resulting RF beatnote, which is recorded on an electrical spectrum analyzer (ESA). The noise of the RF signal generator driving the EO-comb is also analzed directly by the ESA.
			PC $-$ polarization controller; CIR $-$ optical circulator; OSA $-$ optical spectrum analyzer.} 
	\label{fig4}
	\end{figure*}

	\paragraph*{\textbf{Soliton generation with pulsed pumping.}}
For the generation of solitons, we employ optical pumping with pulses~\cite{andersonPhotonicChipbasedResonant2021, brasch_nonlinear_2019}, which has the benefit of reducing thermal effects in comparison to c.w.~operation at the same peak power level. The experimental apparatus is illustrated schematically in Fig.~\ref{fig3}a.
Input pulses of 1.2 ps duration are produced by an electro-optic (EO) comb generator comprising a combination of amplitude and phase modulators, dispersion compensation, and an erbium-doped fiber amplifier (see Supplementary Information).
An additional phase modulator driven by a vector network analyzer (VNA) is used to introduce a small modulation to the pump for probing of the cavity resonance during comb generation~\cite{guo_universal_2017}. The output of the device is measured either in transmission or, as shown in Fig.~\ref{fig3}a, in reflection through an optical circulator, using an optical spectrum analyzer.
A photodiode is used to record a trace of the generated transmitted or reflected light during the frequency scan of the laser and to track the resonance shape.
Because the repetition rate of the EO-comb source is limited for technical reasons to 18~GHz, we pump at 
$f_\mathrm{rep}~$=~$\mathrm{FSR}/4$~=~13.979~GHz. 
Solitons can be generated with average input power in the bus waveguide as low as 23.6~mW.  However, we use 70 -- 80~mW for studying the soliton characteristics, as this  power level is high enough to generate stable, long-lived solitons but low enough to avoid the irreversible damage observed with high-power c.w. pumping (see Supplementary Information). 

The reflected and transmitted spectra observed when pumping with the EO comb centered on the resonance shown in the inset in Fig.~\ref{fig2}c 
are displayed in Fig.~\ref{fig3}(e) and (f).
The superimposed orange spectra  
collected with the pump off resonance are essentially those of the pump pulse. 
Note that in reflection the pump is suppressed by $\sim$10~dB
in comparison to the transmitted spectrum.
Interference fringes observed in the intensity  of the comb spectra are due to reflections between chip facets and the PCRs.

Considering both the transmitted and reflected light, we calculate a conversion efficiency of 2.2\%, defined as the ratio between the power of the soliton lines and the power of the pump (8.9\% taking into account that pumping occurs every fourth pulse).
The detected soliton spectrum contains $\sim$450 lines and spans over 200~nm, or 25~THz, 
covering the reflection bandwidth of the PCRs. Fitting the soliton spectrum with a hyperbolic-secant-squared profile, we infer a full width at half maximum of 3.0~THz, corresponding to a soliton duration of 60~fs.   

In addition to the main grid of soliton lines, there is a weaker set of lines with two local spectral enhancements near 187~THz and 197~THz, most evident in the reflected spectrum. The weaker lines  have a fixed offset of $+\mathrm{FSR}/4$ relative to the main soliton comb but the same mutual spacing of the cavity $\mathrm{FSR}$ (Fig.~\ref{fig3}e insets). 
This indicates that they originate from four-wave-mixing with a phase matching condition between residual EO-comb harmonics and the linear resonances of the PC-FP cavity.  A more detailed explanation supported by numerical simulations is presented in the Supplementary Information.  

In order to study the soliton behavior as a function of detuning from the cold cavity resonance, we apply a method of non-destructive soliton probing~\cite{guo_universal_2017}. 
Sidebands on the pump created by the phase modulator result in an amplitude modulation of the comb generated in the cavity.  Sweeping the modulation frequency with the VNA, we record the response of the cavity as detected by a photodiode and demodulated by the VNA.
In the modulation instability (MI) state (Fig.~\ref{fig3}c), there is one broad peak assigned to the cavity resonance. In the soliton regime, the VNA response has two peaks corresponding to different eigenfrequencies of the field in the quasi-c.w.~($ \mathcal{C} $-resonance) and soliton states ($ \mathcal{S} $-resonance)~\cite{guo_universal_2017}. The frequency of the 
$ \mathcal{C} $-resonance corresponds to the effective detuning from the cold cavity resonance. For 78.6 mW of on-chip input power, the maximum estimated detuning is 3.57 GHz.

Fine tuning of the EO comb repetition frequency is essential to create the proper conditions for the existence of solitons. 
A mismatch of the repetition rate with respect to the cavity FSR at the driving pulse center wavelength leads to a shift in the center frequency of the soliton, which, because of dispersion, changes the soliton's repetition rate~\cite{obrzud_temporal_2017}.
There is a limit however to this mechanism for maintaining the lock of the soliton to the driving pulse~\cite{brasch_nonlinear_2019}.
We measure the permissible deviation of the repetition rate of the EO-comb from one quarter of the 
cavity FSR by recording the generated light with a photodiode during the laser scan for a range of
repetition rates (Fig.~\ref{fig3} (b)). 
We observe a single soliton step over a desynchronization range of 1.3~MHz with a maximum step length of 1.7 GHz. 
The size of this desynchronization range depends nonlinearly with the driving strength and detuning~\cite{erkintalo_phase_2022}, but in general scales positively with the resonator GVD and the inverse of the driving pulse duration. 

	\paragraph*{\textbf{Noise measurement.}}
	The magnitude and source of the soliton's frequency noise are determined by beating individual soliton comb lines reflected from a Bragg-grating notch filter with a low-frequency-noise laser in a heterodyne detection scheme. We analyze the resulting beatnote with an electrical spectrum analyzer (Fig.~\ref{fig4}d).  
	We find that the closer the comb line is to the EO-comb pump frequency, 
	the larger the signal-to-noise ratio (Fig.~\ref{fig4}a, b). 
	In Fig.~\ref{fig4}c, the frequency noise of the EO comb  as well as the selected soliton lines are plotted as a function of the frequency $f$ of the noise.
	The frequency noise of each line $S_{\nu}^{(\mu)}(f)$, where $\mu$ is the mode index, is compared to the frequency noise of the EO comb.  The latter is determined by the frequency noise of the radio-frequency (RF) signal generator used to produce the EO comb.  
	It can be seen that the noise scales with mode index as 
	$S_{\nu}^{(\mu)}(f)=16\mu^2 S_{\mathrm{RF}}(f)$, where the factor of 16 takes into account that $f_\mathrm{rep}=\mathrm{FSR}/4$,  as expected for 
	the soliton being locked to the drive pulse~\cite{brasch_nonlinear_2019}.  The frequency noise of each soliton line  is thus attributed to the noise from the RF generator.

	\subsection*{Discussion}\label{sec12}
	In this work, we demonstrated soliton generation in an integrated PC-FP microresonator made of highly nonlinear GaP. We introduced 
	chirped mirrors, routinely employed in ultrafast laser technology, to the field of DKS generation in microresonators. The net dispersion of the cavity is effectively decoupled from its length and thus the FSR, as the PCRs can be designed to compensate for the dispersion of the central waveguide section to achieve, in principle, any arbitrary dispersion profile over a given frequency range for any cavity roundtrip time. 
	Indeed, in addition to the usual degrees of freedom of width and thickness, the PCRs have many geometric parameters that can be varied.  
	PCRs can be realized with holes of various shapes, external crenulations, or even a series of pillars, each of which can be varied in size, pitch, and shape (\textit{e.g.}, circular, elliptical, or rectangular). 
	One can even combine PCRs with two different dispersion profiles. 
	Comb generation thus becomes possible in wavelength ranges that would otherwise be inaccessible in ring-type resonators, for example, due to strong normal  material dispersion.
	
	The exceptional optical properties afforded by GaP have allowed us to unlock the potential 
	of integrated PC-FP resonators to generate broadband DKSs.  
	Our design provides a linear bandgap evolution for anomalous dispersion as well as an adiabatic transition between the waveguide and the PCRs. Due to the large refractive index of GaP, chirped PCRs were realized with high reflectivity over about 20~THz with only 15 unit cells, while simultaneously achieving a record-high intrinsic quality factor for a GaP resonator of $1.2\times10^6$.  	
	DKS generation with a pulsed pump required only 23.6~mW average input power in our devices, and at slightly higher power, highly stable solitons with a duration of 60 fs were observed, matching the 200-nm bandwidth of the PCRs. Synchronization of the sub-harmonic driving pulse train can deviate over a range of $\sim$1.3~MHz, as the strong nonlinearity of GaP compensates the dispersion of the cavity.
		
	GaP presents numerous opportunities to further exploit the PC-FP concept.
	Thanks to its high index of refraction, high reflectivity is achieved with a small number of unit cells, providing design freedom to tailor group delay over a broad range of frequencies.  Taken together with its wide transparency window (24 THz - 546 THz), we estimate that chirped mirrors for octave-spanning combs are attainable.  A similar design would not be achievable with low-index materials like Si$_3$N$_4$, as extremely long PCRs would be required to cover the broad frequency range with sufficient reflectivity while simultaneously adapting the relative group delay.
	GaP PC-FP resonators could also be leveraged for zero-dispersion solitons~\cite{anderson_zero_2022,xue_dispersion-less_2023, zhang_microresonator_2022} or spectral tailoring~\cite{lucas_tailoring_2022}.
	Finally, the non-zero $\chi^{(2)}$ nonlinearity of GaP opens other possibilities, such as optical harmonic generation~\cite{wolf_quasi-phase-matched_2018}, integrated optical parametric oscillators~\cite{bruch_-chip_2019}, second-harmonic comb generation \cite{wilson_integrated_2020}, and spontaneous parametric down-conversion and generation of half-harmonic combs~\cite{amiune_mid-infrared_2023}.
	
	\subsection*{Methods}
	\paragraph*{\textbf{Resonator dispersion simulation.}}
	The overall dispersion is simulated by dividing the resonator into three sections: straight waveguide, PCRs, and tapered waveguides between the straight waveguide and the PCRs. The effective refractive index $n_\mathrm{eff}$ of the straight waveguide is simulated in two dimensions using COMSOL Multiphysics\textsuperscript{\textregistered} as an eigenfrequency solver. The propagation constant in the straight waveguide is then calculated as 
	\begin{equation}
	\beta=\frac{\partial }{\partial \omega}\left(\frac{n_\mathrm{eff}(\omega)\omega}{c}\right). 
	\end{equation}
	The PCRs are instead simulated with the finite-difference time-domain (FDTD) 3D Electromagnetic Simulator from Lumerical Inc. The group delay of a PCR is calculated from the phase of the reflectivity ($S_{11}$ parameter of the scattering matrix) as $\tau_{\mathrm{g,PCR}} = \frac{\partial  \arg(S_{11}(\omega))}{\partial \omega}$. This contribution to the total group delay takes into account the roundtrip phase shift of the PCR. 
	In an analogous manner, we calculate the group delay of the tapered section $\tau_{\mathrm{g,taper}}$ from the phase shift  of the transmission $S_{21}$.
	
	The sum of the group delays contributed by the individual sections then gives the total group delay for one round trip of the resonator, 
	$GD_{\mathrm{tot}} = 2L_{\mathrm{wg}}\beta + 2\tau_{\mathrm{g,PCR}} + 4\tau_{\mathrm{g,taper}}$.
	The group velocity dispersion of the resonator is then $GVD = \frac{1}{2L}\frac{\partial GD_{\mathrm{tot}}}{\partial \omega}$. For simplicity, we assume a frequency-independent resonator length $L$ equal to the sum of the waveguide length $L_{\mathrm{wg}}$ and two tapers, plus 10~$\mu$m per PCR to account for an average penetration depth into the mirrors.

	\subsection*{Author Contributions}
	A.N. designed the devices. A.N. and C.M. fabricated the sample. A.D., A.N., N.K., M.H.A. and J.R. performed the experiments and data analysis. A.D. and M.H.A. performed numerical simulations of soliton formation. P.S. and T.J.K. conceived and supervised the work.

	\subsection*{Funding}
	This work was supported by the European Union Horizon 2020 Programme for Research and Innovation under grant agreement No. 812818 (Marie Curie ETN MICROCOMB), by the Swiss National Science Foundation (SNF) under grant number 192293, and by the Air Force Office of Scientific Research under award number FA9550-19-1-0250. J. R. acknowledges funding from the SNF Ambizione grant No. 201923 (UNPIC).
	
    \subsection*{Acknowledgments}
     All samples were fabricated at the Binnig and Rohrer Nanotechnology Center (BRNC) at IBM Research Europe, Zurich. We thank Marilyne Sousa, David Indolese, Ute Drechsler and Daniele Caimi for their support in the substrate preparation and characterization.
    
	\subsection*{Disclosures}
	The authors declare no conflicts of interest.
	
	\subsection*{Data Availability}
	The code, data and micrographs used to produce the plots in this work will be released on the repository Zenodo upon publication of this preprint.

	\bibliography{references2.bib,references.bib}

\end{document}


\title{Supplementary Information to: Soliton Microcomb Generation in a III-V Photonic Crystal Cavity}
	
\author{Alberto Nardi}
\thanks{These authors contributed equally to this work.}
\affiliation{IBM Research Europe, Zurich, S\"{a}umerstrasse 4, R\"{u}schlikon, CH-8803 Switzerland}
\affiliation{Institute of Physics, Swiss Federal Institute of Technology Lausanne (EPFL), CH-1015 Lausanne, Switzerland}

\author{Alisa Davydova}
\thanks{These authors contributed equally to this work.}
\affiliation{Institute of Physics, Swiss Federal Institute of Technology Lausanne (EPFL), CH-1015 Lausanne, Switzerland}
\affiliation{Center of Quantum Science and Engineering (EPFL), CH-1015 Lausanne, Switzerland}

\author{Nikolai Kuznetsov}
\thanks{These authors contributed equally to this work.}
\affiliation{Institute of Physics, Swiss Federal Institute of Technology Lausanne (EPFL), CH-1015 Lausanne, Switzerland}
\affiliation{Center of Quantum Science and Engineering (EPFL), CH-1015 Lausanne, Switzerland}

\author{Miles H. Anderson}
\affiliation{Institute of Physics, Swiss Federal Institute of Technology Lausanne (EPFL), CH-1015 Lausanne, Switzerland}
\affiliation{Center of Quantum Science and Engineering (EPFL), CH-1015 Lausanne, Switzerland}

\author{Charles M\"{o}hl}
\affiliation{IBM Research Europe, Zurich, S\"{a}umerstrasse 4, R\"{u}schlikon, CH-8803 Switzerland}

\author{Johann Riemensberger}
\affiliation{Institute of Physics, Swiss Federal Institute of Technology Lausanne (EPFL), CH-1015 Lausanne, Switzerland}
\affiliation{Center of Quantum Science and Engineering (EPFL), CH-1015 Lausanne, Switzerland}

\author{Paul Seidler}
\email[]{pfs@zurich.ibm.com}
\affiliation{IBM Research Europe, Zurich, S\"{a}umerstrasse 4, R\"{u}schlikon, CH-8803 Switzerland}	

\author{Tobias J. Kippenberg}
\email[]{tobias.kippenberg@epfl.ch}
\affiliation{Institute of Physics, Swiss Federal Institute of Technology Lausanne (EPFL), CH-1015 Lausanne, Switzerland}
\affiliation{Center of Quantum Science and Engineering (EPFL), CH-1015 Lausanne, Switzerland}
\maketitle

\setcounter{figure}{0}
\renewcommand{\thefigure}{S\arabic{figure}}
\renewcommand{\theequation}{S\arabic{equation}}

\subsection*{Dispersion engineering of the photonic-crystal Fabry-P\'{e}rot resonator}
Generation of frequency combs and dissipative Kerr solitons in a microresonator depends on the integrated dispersion of the resonator~\cite{kippenberg_dissipative_2018}
\begin{equation}
	D_{\mathrm{int}} = \omega_{\mu}-(\omega_0+D_1\mu) = \sum_{k=2}^{\infty}D_k\frac{\mu^k}{k!},
\end{equation}
where $\omega_0/2\pi$ is the frequency of the pumped resonance (mode index $\mu=0$), $\omega_\mu/2\pi$ is the frequency of the mode with index $\mu$, D$_1/2\pi$ is the free spectral range (FSR) at the pump frequency, and D$_k$ is the k-th order dispersion parameter.
\begin{figure*}[h!]
	\centering
	\includegraphics{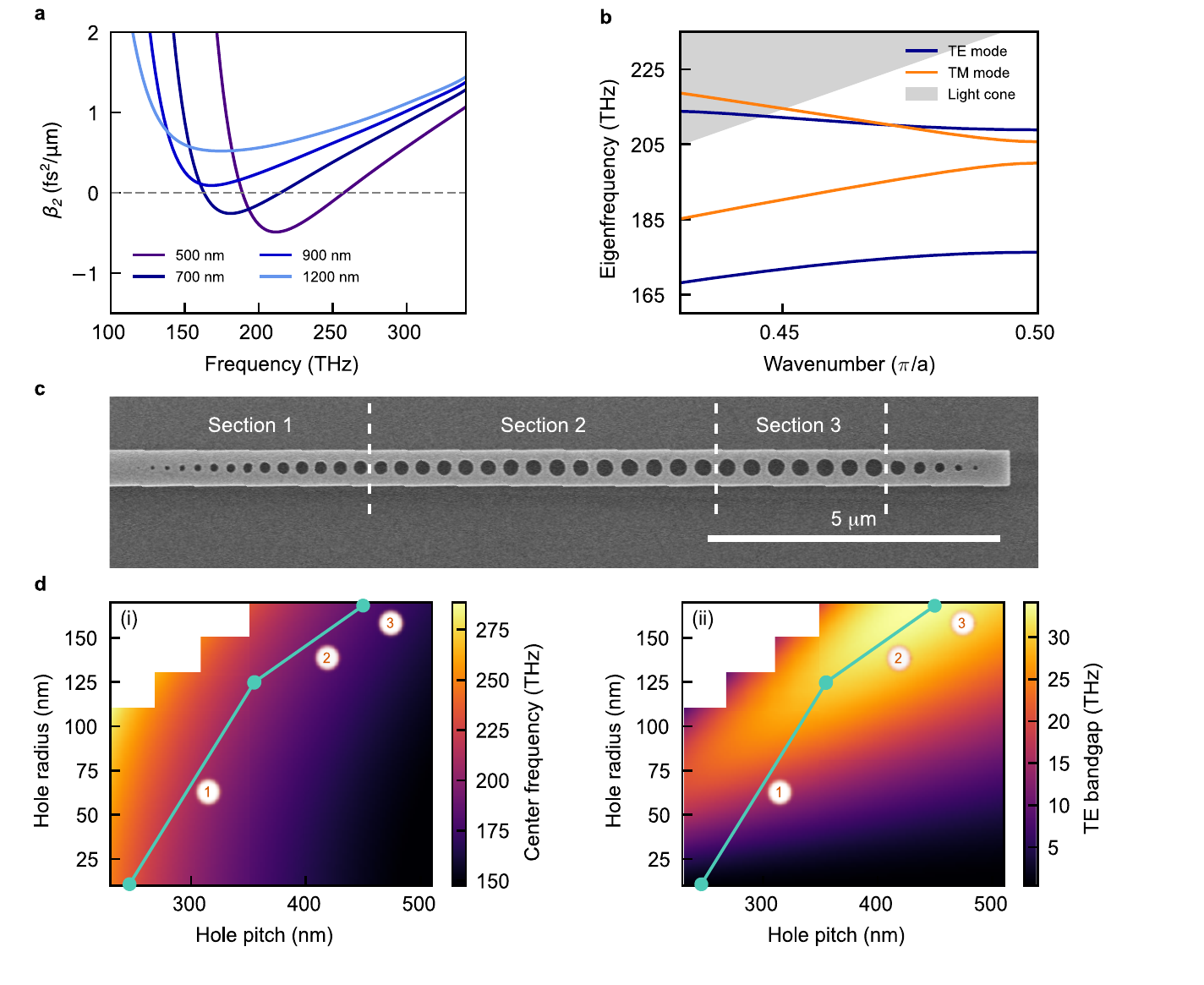}
	\caption{\textbf{Dispersion engineering of straight waveguides and PCRs.} (a) Simulated group velocity dispersion, $\beta_2$, of SiO$_2$-cladded GaP waveguides with a thickness of 300 nm and various widths. A width of 900 nm provides normal dispersion over the entire frequency range with near-zero dispersion around the pump frequency. (b) Photonic band structure near the edge of the first Brillouin zone of a GaP beam (thickness~=~300~nm, width~=~700~nm) with periodic circular holes (hole pitch~=~407~nm, hole radius~=~143~nm) cladded in SiO$_2$.  Only the first two transverse electric (TE) and transverse magnetic (TM) modes are shown. The TE bandgap is maximized with these dimensions and is centered around a frequency of 192.6~THz. (c) Scanning electron microscopy image of a fabricated PCR with the different sections marked as reported in the main text. (d)~Pairwise parameter simulations of the (i) bandgap center frequency and (ii) bandgap as a function of hole pitch and hole radius of a one-dimensional SiO$_2$-cladded GaP photonic crystal with thickness of 300~nm and a width of 700~nm. The segments 1 and 2, as well as the point 3, indicate the geometry of the unit cells for the three different sections of the PCRs. A linear parametrization has been used for the evolution of the unit cell along the segments 1 and 2.
	}\label{fig:SI_Fig_1}
\end{figure*}
To have efficient four-wave mixing (FWM) for generation of bright solitions, the net resonator dispersion should be anomalous and small. We design the waveguide between the photonic crystal resonators (PCRs) to have normal group-velocity dispersion ($\beta_2$~>~0), which for a SiO$_2$-cladded GaP waveguide  with a thickness of 300 nm leads to a choice of 900~nm for the width of the waveguide (see Fig. \ref{fig:SI_Fig_1}(a)). 

Hole-based PCRs behave like distributed Bragg mirrors yielding cavities with a transverse-electric (TE) fundamental mode family. To achieve high reflectivity, the TE bandgap should be maximized (Fig. \ref{fig:SI_Fig_1}(b)).
The mirror design follows a specific parameter evolution of the unit cell dimensions  (varying the hole radius and pitch, while keeping width and thickness constant) for each PCR section (Fig. \ref{fig:SI_Fig_1}(c) and (d)).
An adiabatic transition between the waveguide and PCR is ensured by linearly sweeping the bandgap from 
0.95~THz at a center frequency of 241~THz (hole radius~=~10~nm, hole pitch~=~240~nm) to 31.2~THz at a center frequency of 203~THz (hole radius~=~130~nm, hole pitch~=~370~nm) along segment 1. Then, the bandgap center frequency is linearly swept along segment 2 to achieve high reflectivity between the center frequencies of 203~THz and 176~THz (hole radius~=~170~nm, hole pitch~=~480~nm), while keeping the bandgap constant at about 32~THz. 
A section with the dimensions fixed at those reached at point 3 is added at the end of the mirror to ensure high reflectivity for the lowest frequencies.
The waveguide width was kept constant at 700~nm over the entire PCR.  A narrower width would maximize  the bandgap at higher frequencies.
Our choice of waveguide width provides a sufficiently large bandgap over the entire frequency range while keeping  the losses at longer wavelengths low.
Adjustment of the waveguide width becomes crucial whenever high reflectivity is needed over larger frequency ranges (>~50~THz).

\begin{figure}[h!]
	\centering
	\vspace{0pt}
	\includegraphics[width=0.7\textwidth]{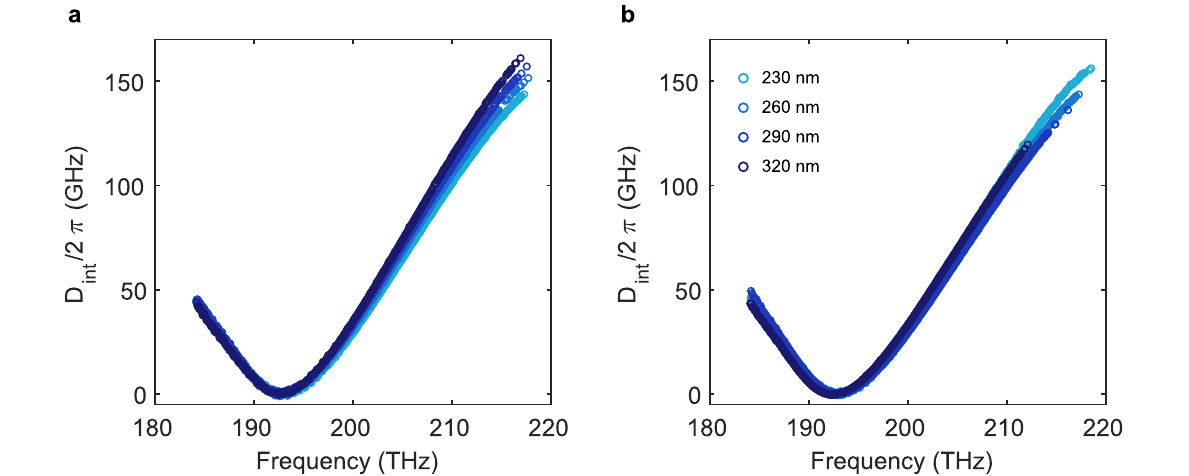}
	\caption{\textbf{Reproducibility of dispersion profile and dependence on the directional coupler.} Devices with a FSR~=~90.1~GHz were compared (waveguide cross-section 900$\times$300~nm$^2$, effective cavity length $L$ = 500 $\mu$m, directional coupler length 25$\mu$m). (a) Measured integrated dispersion of four different device replicas with a gap of 260~nm between the resonator and bus waveguide in the directional coupler. 
	(b) Measured integrated dispersion of devices with various bus-resonator gaps in the directional coupler. 
	}\label{fig:SI_Fig_1bis}
\end{figure}

The fabricated devices exhibit reproducible dispersion profiles, as is evident from the comparison of exact device replicas in Fig. \ref{fig:SI_Fig_1bis}a. We also confirm that the size of the gap between the bus waveguide and the resonator in the directional coupler has no impact on the overall resonator dispersion (Fig. \ref{fig:SI_Fig_1bis}a), indicating that the presence of the directional coupler has negligible effect on the dispersion profile.

\subsection*{Estimation of waveguide propagation losses}

The free-spectral range of a photonic-crystal Fabry-P\'{e}rot (PC-FP) resonator is
\begin{equation}
	\Delta\nu_\mathrm{FSR} =\frac{1}{T_\mathrm{R}} = \frac{c}{2n_\mathrm{g} L},
\end{equation}
where $T_\mathrm{R}$ is the resonator roundtrip time, $c$ is the speed of light, $n_\mathrm{g}$ is the effective group index of the resonator, and $L$ is the effective cavity length; the factor of $2$ accounts for the roundtrip resonator length.  Although $L$ depends on frequency, we assume $L = L_\mathrm{wg} + 2L_\mathrm{m}$, where $L_\mathrm{m}$ is the length of the taper connecting the waveguide section with the PCR plus 10 $\mu$m to account for some penetration into the PCRs.  In other words, we approximate $L$ and $L_\mathrm{m}$ as being constant.

The intrinsic loss rate $\kappa_0$ is given by the sum over several loss channels in the resonator:
\begin{equation}
	\kappa_0 = \kappa_\mathrm{m} + \kappa_\mathrm{p}^\mathrm{m} + \kappa_\mathrm{p}^\mathrm{wg},
\end{equation}
where $\kappa_\mathrm{m}$ describes the scattering losses in the PCRs, while $\kappa_\mathrm{p}^\mathrm{m}$ and $\kappa_\mathrm{p}^\mathrm{wg}$  represent the propagation losses at the ends of the resonator (tapers and PCRs) and in the waveguide, respectively. 
The scattering loss in the PCRs is given by a splitting ratio per roundtrip $m$:
\begin{equation}
	\kappa_{m} = \frac{2m}{T_\mathrm{R}} = m \frac{c}{n_\mathrm{g} L}.
\end{equation}
The contributions from the propagation losses depend on the corresponding attenuation coefficients per unit length, $\alpha_\mathrm{wg}$ and $\alpha_\mathrm{m}$, as 
\begin{equation}
	\kappa_\mathrm{p}^\mathrm{m} = \frac{c}{n_\mathrm{g}^\mathrm{m}}\alpha_\mathrm{m} \quad\mathrm{and}\quad \kappa_\mathrm{p}^\mathrm{wg} = \frac{c}{n_\mathrm{g}^\mathrm{m}}\alpha_\mathrm{wg},
\end{equation}
where $n_\mathrm{g}^\mathrm{m}$ and $n_\mathrm{g}^\mathrm{wg}$ are the effective group indices for the taper-PCR and waveguide portions of the resonator, respectively.

We can then write
\begin{equation}
	\kappa_0 = \frac{c}{2n_\mathrm{g} L}\left(2m + \frac{n_\mathrm{g}}{n_\mathrm{g}^\mathrm{m}}4L_\mathrm{m}\alpha_\mathrm{m} + \frac{n_\mathrm{g}}{n_\mathrm{g}^\mathrm{wg}}2L_\mathrm{wg}\alpha_\mathrm{wg}\right).
\end{equation}
Note that this model assumes that the overall propagation loss per roundtrip is small.

Substituting  $L - 2L_\mathrm{m}$ for $L_\mathrm{wg}$ gives a linear dependence of $\kappa_0$ on $1/L$:
\begin{equation}
	\kappa_0 = \frac{c}{n_\mathrm{g}}\left[m + 2L_\mathrm{m} \left(\frac{n_\mathrm{g}}{n_\mathrm{g}^\mathrm{m}}\alpha_\mathrm{m} - \frac{n_\mathrm{g}}{n_\mathrm{g}^\mathrm{wg}}\alpha_\mathrm{wg}\right)\right]\frac{1}{L} + \frac{c}{n_\mathrm{g}^\mathrm{wg}}\alpha_\mathrm{wg}.
\end{equation}
The term in brackets can be viewed as the fraction of light not reflected at the ends of the resonator, $1-r$, giving
\begin{equation}
	\kappa_0 = \frac{c(1-r)}{n_\mathrm{g}L} + \frac{c}{n_\mathrm{g}^\mathrm{wg}}\alpha_\mathrm{wg}.
\end{equation}
The effective group index for the resonator  $n_\mathrm{g}$ is less than 0.5\% lower than the group index of the waveguide section $n_\mathrm{g}^\mathrm{wg}$.  Thus, 
\begin{equation}
	\kappa_0 \approx \frac{c}{n_\mathrm{g}}\left(\frac{1-r}{L} +\alpha_\mathrm{wg}\right).
\end{equation}
 We can therefore determine the propagation loss in the waveguide, $\alpha_\mathrm{wg}$, from the intercept of a plot of $\kappa_0$ versus $1/L$, \textit{i.e.}, for an infinite waveguide length ($L\to\infty$), the mirror and taper contributions to the intrinsic losses become negligible.  

The slope of the line depends on the scattering loss in the PCRs, on the length of the taper-PCR section, on the propagation loss coefficients in the different sections of the resonator, and the relative effective group indices.  As the PCRs and the length of the tapers of the devices compared in Fig. 2b in the main text are identical,  
we attribute the difference in slope for devices with a waveguide width of  900 nm and 1200 nm to different propagation losses in the taper-PCR sections. 

\subsection*{Fabrication process}

To fabricate the PC-FP resonators, we use a sacrificial single-side-polished, [100]-oriented, GaP wafer as a template to grow an epitaxial GaP/Al$_{0.2}$Ga$_{0.8}$P/GaP stack by metal-organic chemical vapor deposition. An initial 100-nm-thick GaP buffer layer facilitates nucleation and growth of the subsequent 200-nm-thick Al$_{0.2}$Ga$_{0.8}$P layer. The thickness of the top GaP device layer depends on device design; in this work, we used a thickness of 300~nm. After epitaxial growth, we bond the wafer onto a Si wafer capped with 2 $\mu$m of thermal dry SiO$_2$. Prior to bonding, the surface of both wafers is coated with 5 nm of Al$_2$O$_3$ deposited by atomic-layer deposition. The bonded wafers are then annealed at 250~$^{\circ}$C.

\begin{figure}[t]
	\centering
	\vspace{0pt}
	\includegraphics{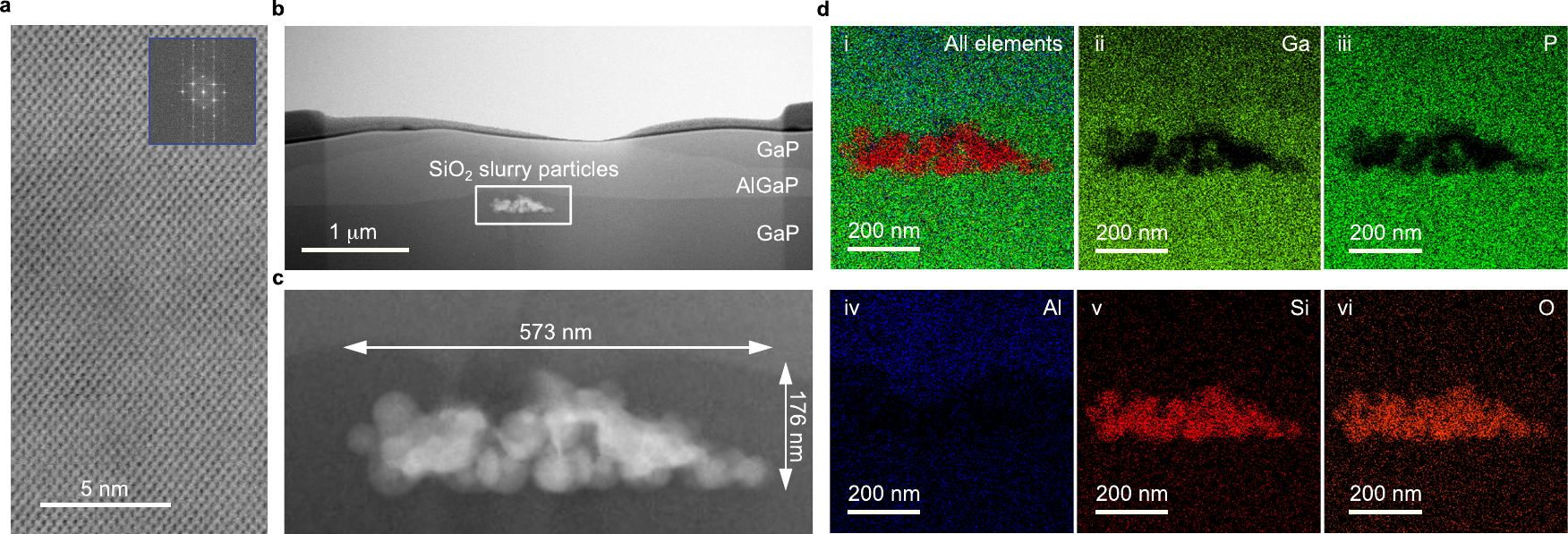}
	\caption{\textbf{Characterization of epitaxial growth by transmission electron microscopy.} (a) Crystalline structure of the 100-nm-thick GaP buffer layer. Inset: Fourier transform showing the crystal lattice of GaP in reciprocal space. No defects are observed. Similar results were also obtained for the Al$_{0.2}$Ga$_{0.8}$P layer except in the region above the agglomerate of silica slurry particles.  (b) Bright field image of the grown material stack in the vicinity of a defect. No interface can be seen between the GaP sacrificial wafer and the GaP buffer layer. Three-dimensional growth is triggered by agglomerations of silica particles left during chemical-mechanical polishing  on the purchased wafers. The thickness of the Al$_{0.2}$Ga$_{0.8}$P layer is not uniform due to different growth rates along distinct crystalline directions. (c) Magnified image of the silica slurry particles. (d)~Energy-dispersive X-ray spectroscopy for elemental analysis carried out with a liquid-nitrogen-free silicon drift detector. (i-vi) Spatial maps of the K$_\alpha$ X-ray emission from various elements, establishing the material composition of the analyzed region.
	}
	\label{fig:SI_Fig_TEM}
\end{figure}

\begin{figure}[b]
	\centering
	\vspace{0pt}
	\includegraphics{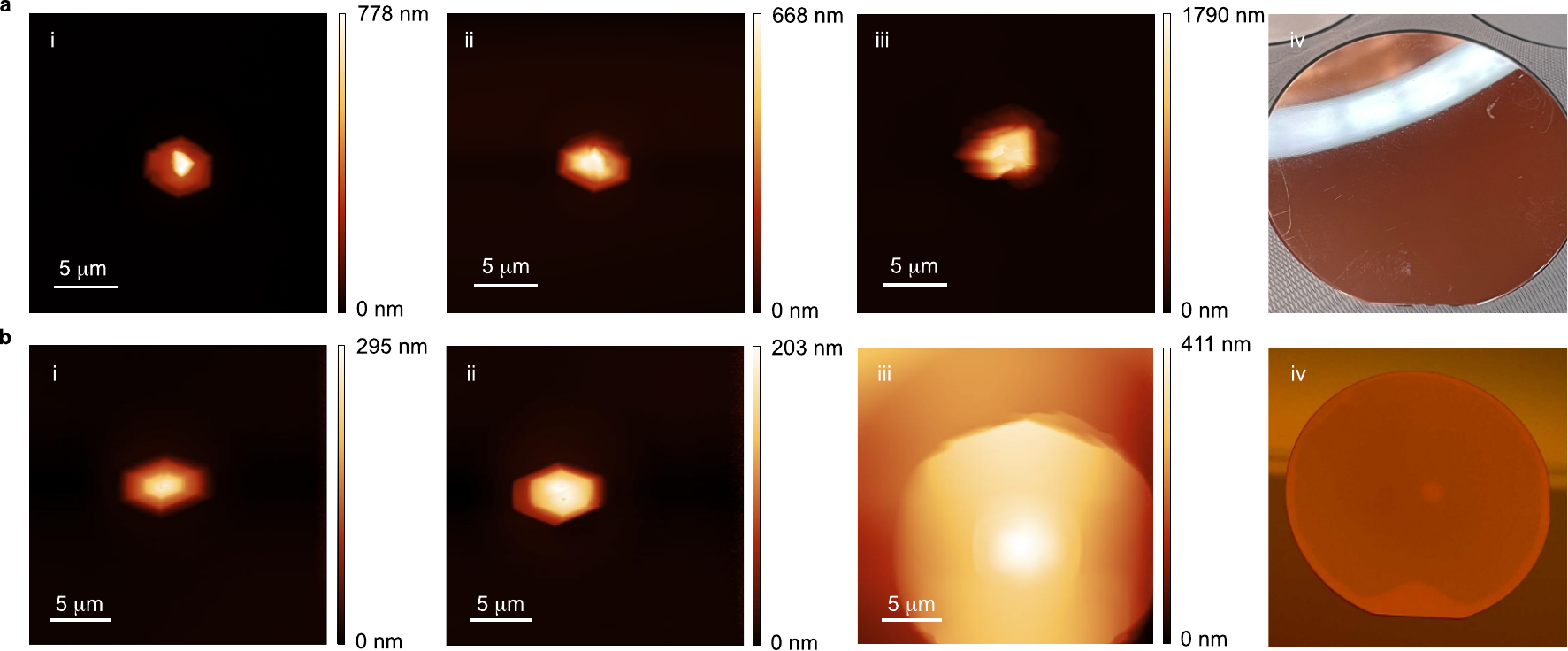}
	\caption{\textbf{Hillock formation during epitaxial growth and bonding results.} 
		Epitaxially grown material stacks consisting of 100~nm GaP/200~nm Al$_{0.2}$Ga$_{0.8}$P/200~nm GaP are compared.  (a) Results without repolishing.  (i-iii) Atomic force microscopy shows hillocks ranging in height from about 700~nm to a few micrometers.  (iv) Image of 2-inch wafer after epitaxial growth. Defects are visible all over the surface, making wafer bonding impossible.  (b) Results with re-polishing. (i-iii) Atomic force microscopy shows that the height of the hillocks has been reduced to 200 -- 400~nm. (iv) Image of 2-inch bonded wafer.  The bonding yield is nearly 100\%, except for a defect in the middle of the wafer.
	}
	\label{fig:SI_Fig_hillocks}
\end{figure}

Despite the good crystalline quality of the epitaxially grown layers (Fig. \ref{fig:SI_Fig_TEM}(a)), surface contamination of the purchased sacrificial GaP wafers can be an issue.  Specifically, slurry particles can produce defects in the device layer that are detrimental to the overall wafer bonding yield. Slurry particles are left on the wafer surface by the chemical-mechanical polishing performed after the wafer is cut out of the GaP ingot. These particles can initiate three-dimensional crystal growth with growth rates that depend on the crystal direction, usually leading to the development of hexagonal hillocks with variable height (Fig. \ref{fig:SI_Fig_hillocks}). We confirmed that the origin of the growth defects is the presence of silica slurry particles using transmission electron microscopy of a lamella prepared containing a hillock (Fig. \ref{fig:SI_Fig_TEM}(b-d)).  

In an attempt at removing the silica particles from the surface, several wet etching processes (\textit{e.g.}, buffered HF, piranha, aqua regia) were tested on the wafers before growth, but no chemistry has been found that successfully removes the particles without damaging the surface, and the further handling often introduces additional particles. 
The only viable solution, in our experience, is re-polishing of the wafer surface by a commercial polishing company. Without re-polishing (Fig. \ref{fig:SI_Fig_hillocks}(a)), hillocks form with a wide distribution of heights, ranging from several hundred nanometers to a few micrometers, and bonding yield is usually around 50\%. Following re-polishing (Fig. \ref{fig:SI_Fig_hillocks}(b)), the hillock density is much lower, and their average height is only a few hundred nanometers, leading to nearly complete bonding of the wafer. 

After bonding, most of the sacrificial GaP wafer is removed by mechanical grinding.  The last 50~$\mu$m are dry etched in a mixture of SF$_6$ and SiCl$_4$ in an inductively coupled-plasma reactive-ion-etching (ICP-RIE) tool. In this process, the Al$_{0.2}$Ga$_{0.8}$P layer acts as an etch stop, protecting the GaP device layer. The Al$_{0.2}$Ga$_{0.8}$P layer is subsequently removed by selective wet etching in concentrated HCl for 4~minutes. 

To pattern devices in the GaP device layer, we employ hydrogen silsesquioxane (HSQ) as a negative resist for electron-beam lithography. The pattern is transferred into the GaP by ICP-RIE using a mixture of BCl$_3$,Cl$_2$, CH$_4$, and H$_2$. 
After removing the HSQ in buffered HF, a photolithographic step is used to expose the inverse tapers at the edge of the chip for fiber coupling. The sample is covered with 6 $\mu$m of AZ4562 photoresist (MicroChemicals GmbH) to protect the devices during etching. The photoresist is exposed to UV light in a mask aligner and developed to remove the resist near the edges of the chip. 
The underlying 
SiO$_2$ layer is etched in an ICP-RIE process based on C$_4$F$_8$ and O$_2$, and about 150~$\mu$m of the silicon substrate are removed by deep reactive-ion etching using SF$_6$ and C$_4$F$_8$ (Bosch process). Finally, the remaining photoresist is stripped in acetone, and any residues are removed in an oxygen plasma asher. The facets of the inverse tapers are now accessible for efficient coupling into the chip with lensed fibers. 

\subsection*{Raw characterization spectra}

\begin{figure}[h]
	\centering
	\vspace{0pt}
	\includegraphics[width=0.85\linewidth]{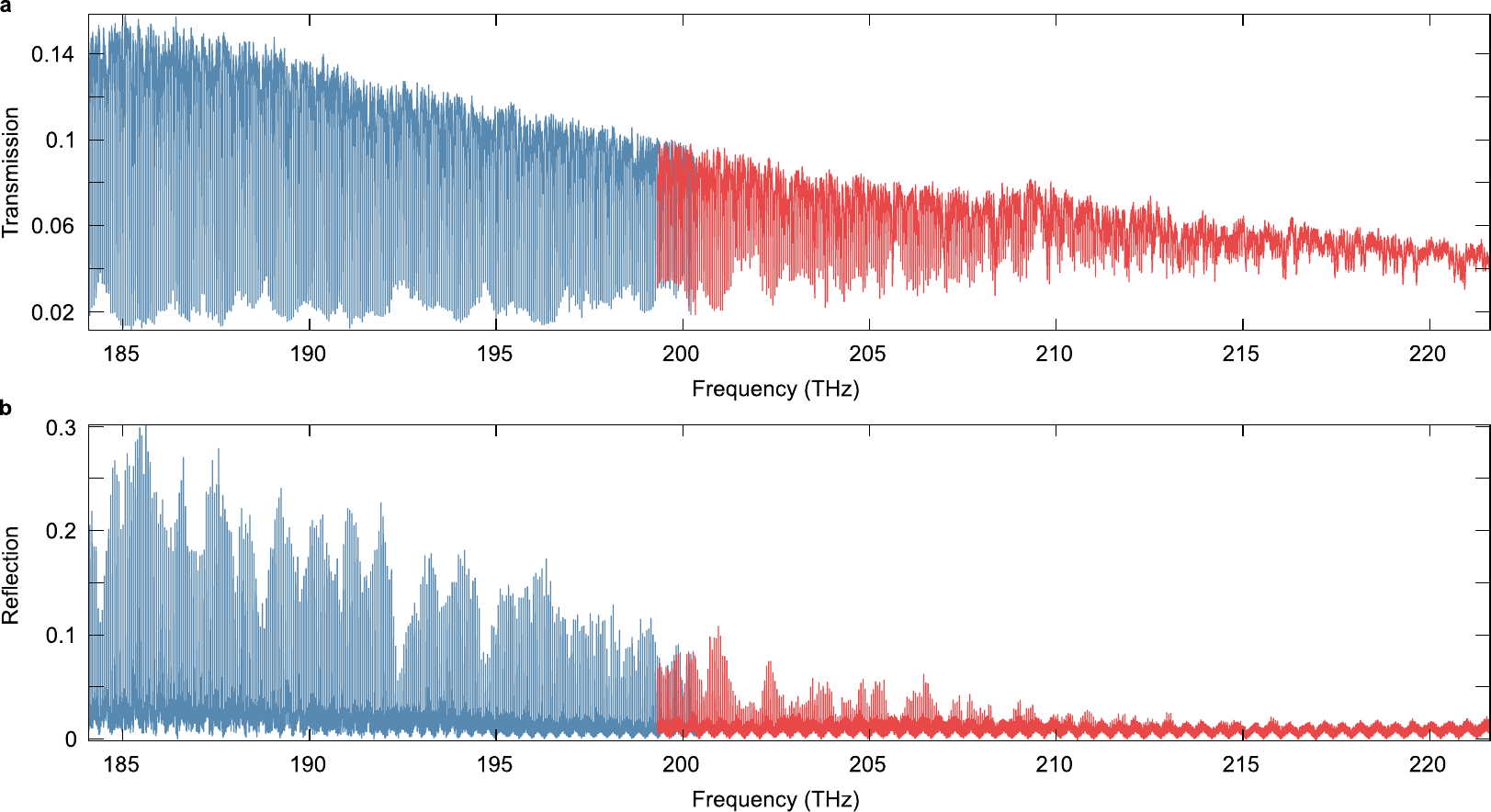}
	\caption{\textbf{Raw characterization spectra.} (a) Transmission and (b) reflection spectra of the device used for soliton generation. The blue and red traces are recorded with two different lasers. 
	}
	\label{fig:SI_Fig_8}
\end{figure}

The raw transmission and reflection spectra for the device measured in the main manuscript are shown in Fig. \ref{fig:SI_Fig_8}(a-b). Two external-cavity diode lasers are used to cover a range of about 40 THz.  The frequency axes of the spectra are calibrated using a separate fiber-laser frequency comb~\cite{delhaye_frequency_2009}. Because light circulates back and forth in the PC-FP resonator, sharp resonances are visible in both transmission and reflection.
Fiber-to-fiber transmission around 193~THz is $\sim$11\%. 
To produce Fig. 2(c) in the main text, the traces from the two lasers are stitched together, and a Savitsky-Golay filter with a width of 100 GHz is applied to remove the interference fringes produced by reflections from the chip facets.

\subsection*{Continuous-wave pumping}

\begin{figure}[h]
		\centering
		\vspace{0pt}
		\includegraphics[width=0.45\linewidth]{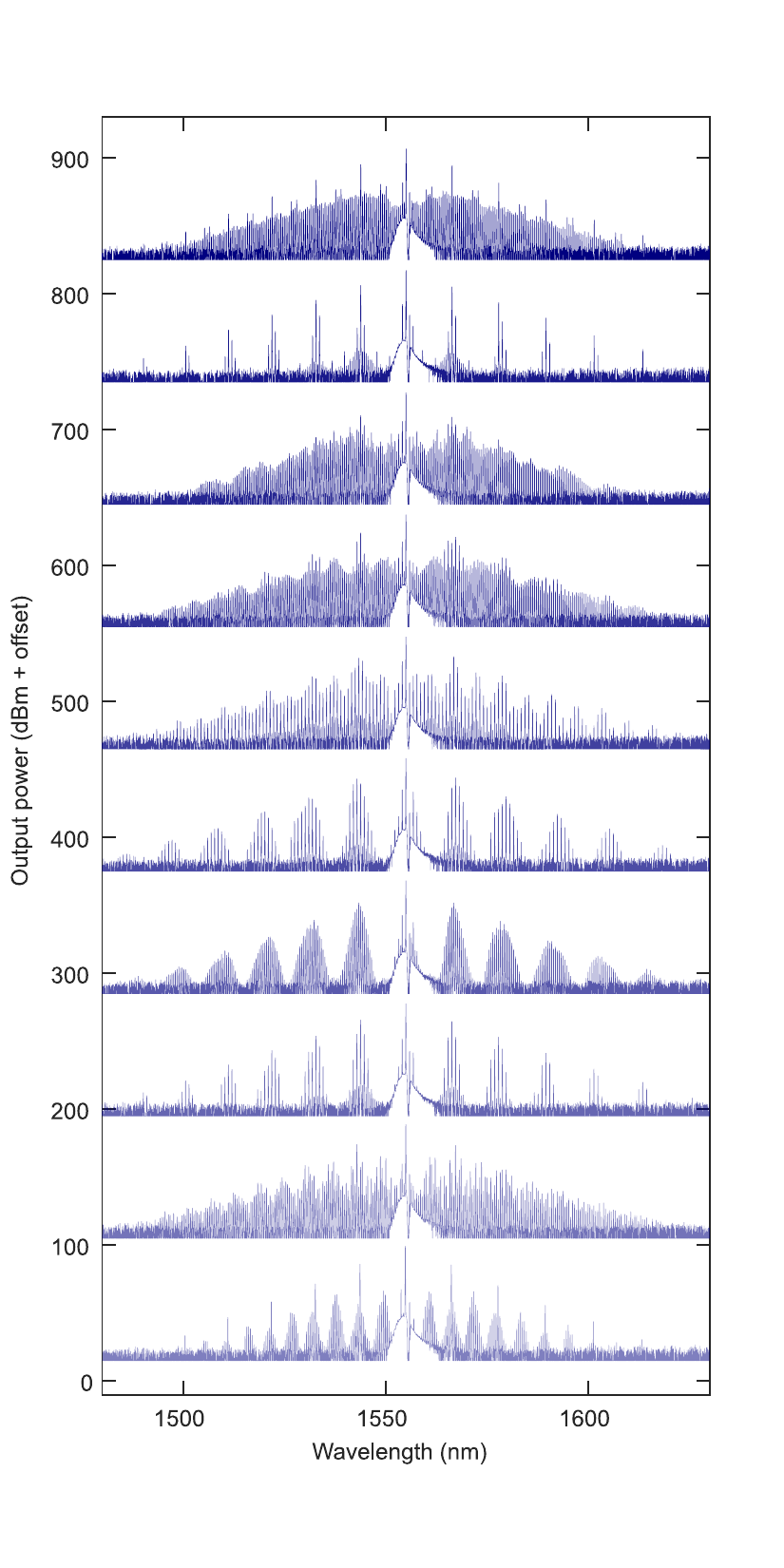}
		\caption{\textbf{Frequency comb evolution with c.w. pumping.} From top to bottom, the laser–cavity detuning is incrementally reduced over 186~pm with the laser power fixed. The frequency comb undergoes abrupt switches between different comb states.}
		\label{fig:SI_Fig_6}
	\end{figure}

Attempts to generate solitons by pumping the PC-FP resonators with a scanning continuous-wave (c.w.) source failed due to the large thermorefractive effect~\cite{moilleDissipativeKerrSolitons2020} in GaP; the nonlinear distortion of the resonances at pump powers above threshold was so great that it was not possible to scan over the entire resonance to reach the soliton state with the maximum accessible laser scanning range of 36~GHz.
In addition, on-chip input power levels above 150~mW caused irreversible damage to the devices, degrading the quality factor of the resonances. The frequency comb evolution observed with input power of 150~mW for a PC-FP resonator with FSR~=~55.916~GHz over a laser detuning range of 186~pm is shown in (Fig. \ref{fig:SI_Fig_6}). The comb lines rearrange several times as the detuning is reduced, alternating between a broad single comb spanning up to 120~nm and smaller secondary combs within localized gain lobes. 
	
\subsection*{Electro-optic comb generation}

\begin{figure}[h!]
	\vspace{0pt}
	\includegraphics{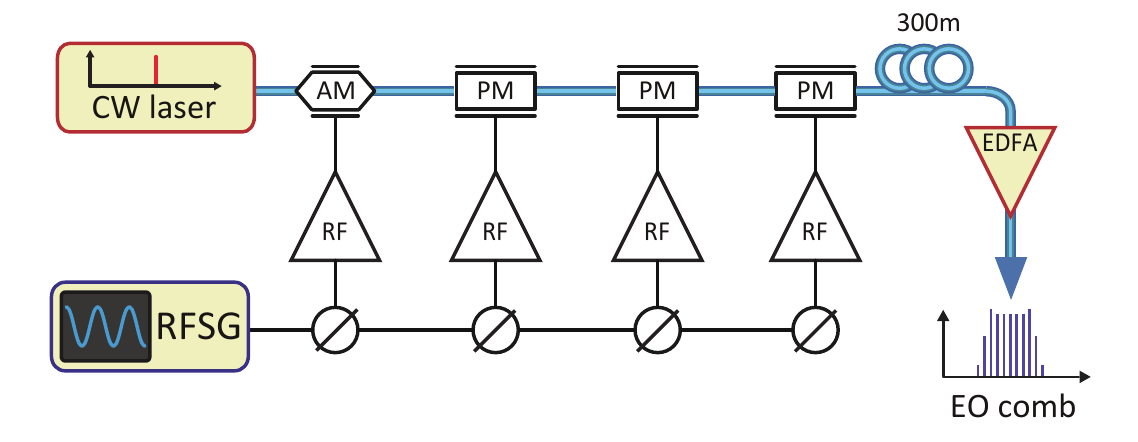}
	\caption{\textbf{Schematic of the apparatus to generate the EO comb:} c.w. laser source; AM $-$ amplitude modulator; PM $-$ phase modulator; RFSG $-$ radio-frequency signal generator; RF $-$ radiofrequency amplifier; EDFA $-$ erbium-doped fiber amplifier.}\label{fig:SI_Fig_3}
\end{figure}

\begin{figure}
	\vspace{0pt}
	\includegraphics{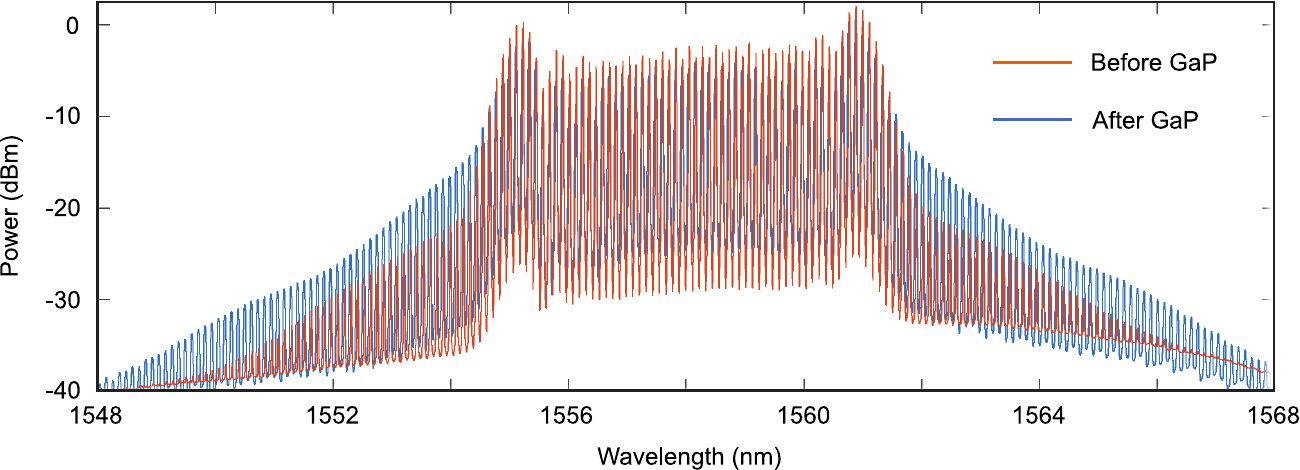}
	\caption{\textbf{Measured spectra of the EO-comb pulses after the amplification.} The spectra are measured before (red) and after (blue) propagating through the GaP sample with average power of 70 mW in the bus waveguide. The spectrum is broadened after passing through the GaP waveguide due to self-phase modulation.}\label{fig:SI_Fig_3.1}
\end{figure}

The pulse used to drive the PC-FP resonators is formed via hybrid electro-optic (EO) modulation of a c.w.~source and subsequent compression in a dispersive medium. Multiple sidebands with a flat spectral envelope are obtained with a combination of  an amplitude mudulator (AM) and three phase modulators (PM)~\cite{kobayashiOpticalPulseCompression1988, obrzud_temporal_2017} (Fig.~\ref{fig:SI_Fig_3}). 
The modulating radio-frequency signal generated with a Rhode \& Schwarz SMB100A RFSG is split into four channels with independent amplifiers to drive each of the modulators. The resulting modulated optical field before compression consists of a multi-carrier signal with a chirped phase. A 300-m-long single-mode fiber (SMF-28) and a 10-m-long dispersion compensation fiber (DCF) were employed to compensate the chirp and compress the signal into a pulse. By adjusting the relative phases of the phase modulators and the voltage offset of the amplitude modulator, the optimal chirp leading to the shortest pulse duration is obtained.  Given that each modulator has an insertion loss of about 3~dB, the output must be amplified with an EDFA to obtain sufficient power for Kerr comb generation. After the pulse compression stage and the EDFA, the resulting pulse has a duration of 1--1.2~ps, depending on the amplification level of the EDFA.
The pulse spectrum broadens even further after traveling through the GaP bus waveguide on the measured sample due to its large nonlinearity (see Fig.~\ref{fig:SI_Fig_3.1}). 

\subsection*{FROG measurements}

\begin{figure}[h!]
	\centering
	\vspace{0pt}
	\includegraphics{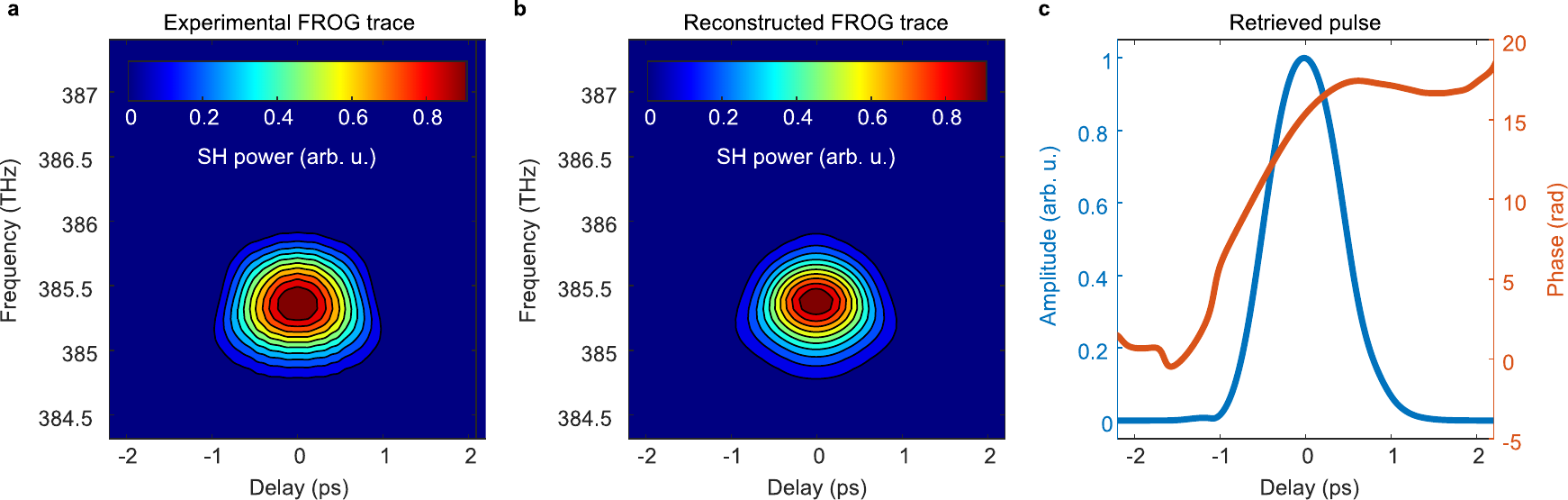}
	\caption{\textbf{Frequency-resolved optical gating measurements of the EO-comb pulse used for soliton generation.} (a) Experimentally measured trace. (b) Reconstructed trace. (c) Amplitude and phase of the pulse retrieved iteratively.}
	\label{fig:SI_Fig_7}
	
\end{figure}

To determine the shape of the EO-comb pulse used during soliton generation, we perform frequency-resolved optical-gating (FROG) measurements~\cite{d_j_kane_characterization_1993, trebino_measuring_1997, delong_frequency-resolved_1994}. 
During the soliton generation process, a portion of the EO-comb pulse is sent to a custom free-space optical FROG setup. The pulse is split in a balanced Michelson interferometer, and a high-precision motorized stage is used in one of the interferometer arms to control the delay between the two spatially parallel output pulses. They are then focused onto the same point on a tilted $\beta$-barium borate crystal. The experiment is operated in the noncollinear regime producing three type-I second-harmonic beams. The intensity of the center pulse, which is present only when the delay between the two pulse copies is smaller than the initial pulse width, is measured with a spectrometer as a function of delay, producing a FROG trace. To maximize the efficiency of second-harmonic generation (SHG), $ \lambda/2$ and $\lambda/4$ plates are employed to adjust the polarization of the input pulse before it enters the Michelson interferometer. To retrieve the pulse shape from the measured FROG trace, we use an algorithm that iteratively makes a prediction of the pulse shape and, using a Frobenius norm, compares it with the experimental trace. The pulse shape is retrieved when the error is minimized after multiple iterations. The experimental FROG trace of the EO-comb pulse is shown in Fig.~\ref{fig:SI_Fig_7}(a), while the reconstructed FROG trace and pulse are shown in Fig.~\ref{fig:SI_Fig_7}(b) and  Fig.~\ref{fig:SI_Fig_7}(c), respectively. As expected, it is a single pulse with a full width at half maximum (FWHM) of 1.0~ps, in good agreement with the simulated pulse duration.

\subsection*{Dissipative soliton simulations with modified Lugiato-Lefever mean-field model}

\begin{figure}[h!]
	\vspace{0pt}
	\includegraphics{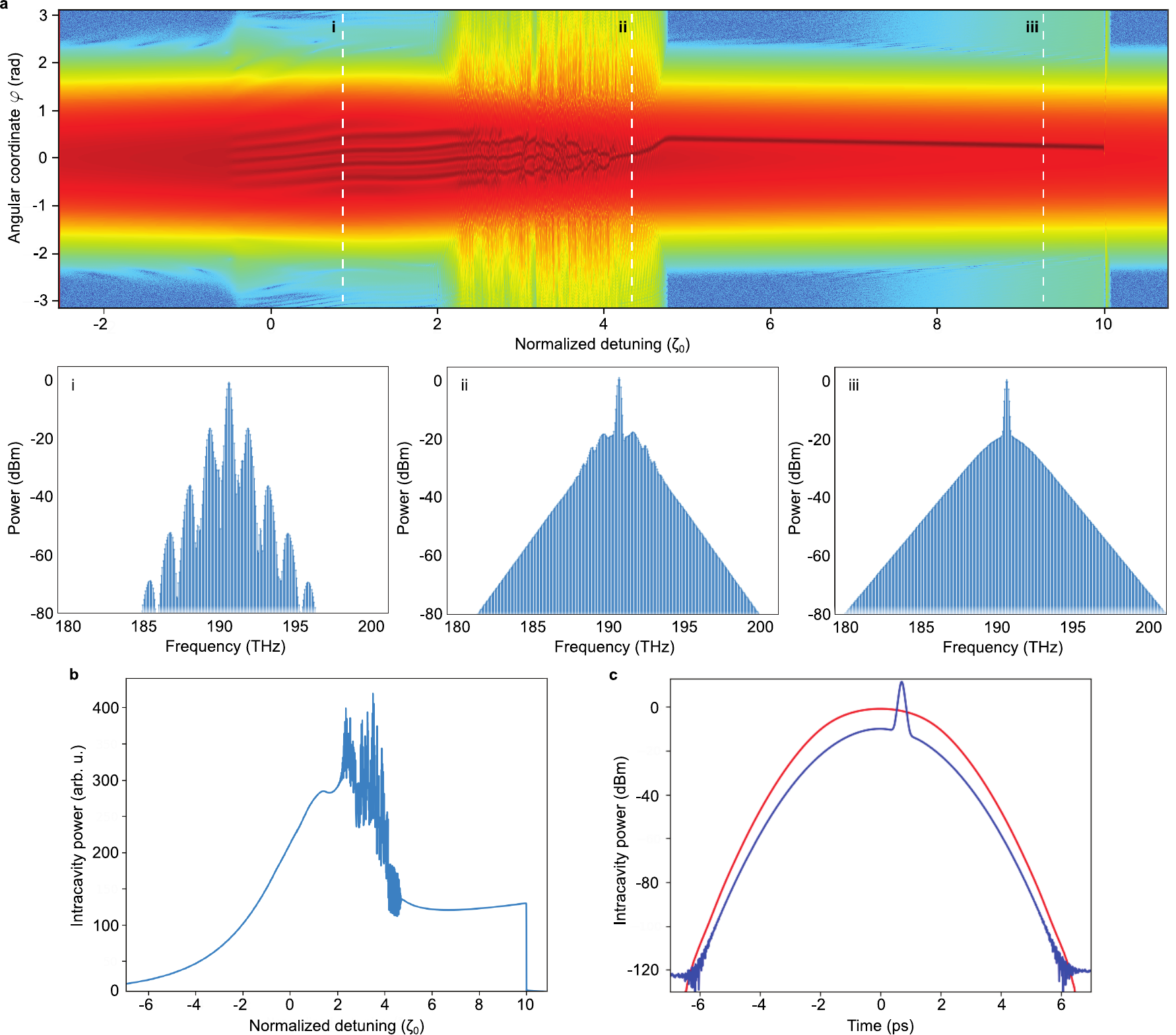}
	\caption{\textbf{Modified-LLE simulations.} (a) Simulation of the time-domain intracavity field in a pulse-driven Fabry-P\'{e}rot microresonator as a function of normalized detuning across the resonance (where $\zeta_0=2\delta\omega/\kappa$). The angular coordinate $\varphi$ is defined as $2\pi \times t / T_{\mathrm{roundtrip}}$, where $t$ is the real time and $T_{\mathrm{roundtrip}}$ is the resonator roundtrip time. The color axis representing energy density is in a logarithmic scale. (i-iii) Output spectra for various detuning slices showing the modulation-instability, breathing, and single-soliton states, respectively. (b) Average intracavity power versus normalized detuning. A single soliton step is clearly visible. (c) Optical power of the input pump pulse (red) and the generated soliton (blue) in the time domain.}\label{fig:SI_Fig_4}
	
\end{figure}
Soliton dynamics in Fabry-P\'{e}rot-type resonators has been shown to be slightly different than in ring-type resonators \cite{cole_theory_2018}. However, the mean-field model can still be used, and the system is described via a modified Lugiato-Lefever equation (LLE).
The modified LLE is derived from the coupled mode equations by adding a cross-phase modulation (XPM) term, $2i\sum_{\mu'}|b_{\mu'}|^2a_{\mu}$, to account for the additional phase shift caused by the counter-propagating field. We make the same approximation as in~\cite{obrzud_temporal_2017}, namely that the counter-propagating field $b_{\mu}$ evolves more slowly than the roundtrip time. Therefore, the XPM term becomes  $2i\sum_{\mu'}|a_{\mu'}|^2a_{\mu}$, and we have
\begin{equation}
\frac{\partial a_{\mu}}{\partial{\tau}}=-(1+i \zeta_{\mu})a_{\mu}+i\sum_{\mu', \mu''}a_{\mu'}a_{\mu''}a^*_{\mu'+\mu''-\mu}+2i\sum_{\mu'}|a_{\mu'}|^2a_{\mu}+f_{\mu},
\end{equation}
where $\zeta_{\mu}= \zeta_0+d_2\mu^2+d_3\mu^3$ represents the dimensionless dispersion and detuning operator. 
Converting back into the rotating spatial frame of the resonator (angular coordinate $\varphi$) over real time $t$ and with practical experimental units, we recall the typical LLE now with the counter-propagating XPM modifier to the detuning
\begin{equation}
\frac{\partial A}{\partial t} =-(\frac{\kappa}{2} +i\delta\omega)A +i\frac{D_2}{2}\frac{\partial^2 A}{\partial\varphi^2} +\frac{D_3}{6}\frac{\partial^3 A}{\partial\varphi^3} +i\Gamma\big[ |A|^2+2\langle |A|^2 \rangle\big] A +\sqrt{\frac{\kappa_\mathrm{ex}D_1 P_\mathrm{in}(\varphi)\delta_{T4}}{2\pi}},
\end{equation}
acting on intracavity field $A(\varphi,t)$ with dimension of $\sqrt{\mathrm{W}}$. Here the nonlinear parameter $\Gamma$ is related to the more known waveguide nonlinearity $\gamma$ by
\begin{equation}
	\Gamma = \frac{\gamma LD_1}{2\pi} = \frac{n_2\omega_0 LD_1}{2\pi cA_\mathrm{eff}}
\end{equation}
$P_\mathrm{in}(\varphi)$ represents the injected driving pulse term, and $\delta_{T4}$ is a conditional parameter that is `on' (equals 1, or else 0) only on every fourth roundtrip of the resonator over the real time $t$, to realistically recreate the experimental pumping conditions.
The dispersion coefficients used in the simulations are taken from the experimental data up to the third order: D$_2$/(2$\pi$)~$=$~3.03~MHz, D$_3$/(2$\pi$)~$=$~2.82~kHz. The pump pulse was approximated as a Gaussian pulse with a width of 1.1~ps and a normalized peak power sufficient for soliton formation, in this case $P_0/P_\mathrm{thresh}=10$. Simulations of soliton formation (Fig. \ref{fig:SI_Fig_4}) predict a soliton spectrum with a FWHM of 3.0~THz, in agreement with our experimental results.

\subsection*{Explanation of subcombs}

\begin{figure}[H]
	\centering
	\includegraphics{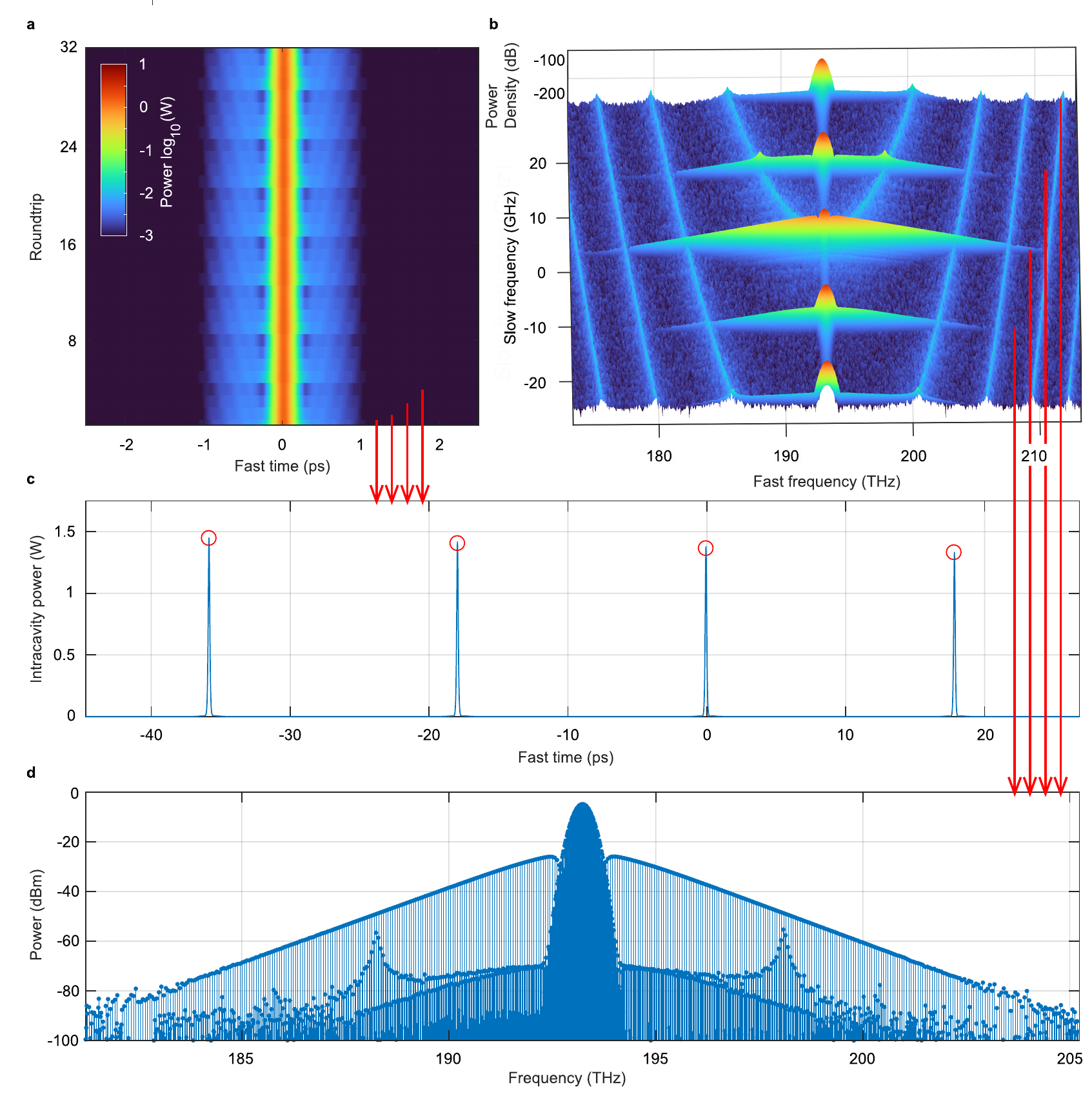}
	\caption{\textbf{Simulation of 1-4 sub-harmonic pulse-driving.} (a) Intracavity optical field for each roundtrip. The field decays in a cycle with a duration of four roundtrips, and a radiating wave can be seen. (b) Two-dimensional Fourier transform of the field in (a), with power spectral-density between each FSR plotted vertically in each column, for each FSR comb index on the $x$-axis. The four comb tranches are phase-locked with the FSR/4 EO comb. (c) Four intracavity fields plotted in sequence, comprising one whole EO-comb period, showing the decay in soliton amplitude over four round trips. (d) The four comb orders from (b) interleaved, separated by FSR/4~=~14 GHz. }
	\label{fig:SI_Fig_SubHarmonicSim}
\end{figure}	

In our experiment, the soliton is not being driven by the pulsed pump every single roundtrip, but instead only every fourth roundtrip, with a pulse having $4^2$ times the energy necessary for the same soliton to exist if it were pumped every roundtrip. This is demonstrated in a simulation of the LLE shown in Fig. \ref{fig:SI_Fig_SubHarmonicSim}. Essentially, within the \emph{mean-field model} (the LLE), the pump is only present 25\% of the time. Because of this, the soliton is perturbed once every four roundtrips and undergoes decay for the other three, as observed in Fig. \ref{fig:SI_Fig_SubHarmonicSim}(a) and (c), forming a saw-tooth-shaped energy pattern over time and radiating weak waves in each direction. In the frequency domain (Fig. \ref{fig:SI_Fig_SubHarmonicSim}(b) and (d)), this causes coupling via FWM, leading to energy transfer from the broad soliton to frequencies existing at $\pm n\mathrm{FSR}/4\approx \pm n \times 14$ GHz \cite{anderson_dissipative_2023}, in phase with the other three unused EO-comb orders. These subcomb lines can be seen broadly at a low level in Fig. 3(e) of the main text, and replicated in the modelling in Fig. \ref{fig:SI_Fig_SubHarmonicSim}(d). In particular, where these off-order comb lines come into resonance with the passive dispersion of the resonance spectrum around $D_\mathrm{int} = \mu^2D_2/2 +\mu^3D_3/6$ (seen in the parabolic curves in Fig. \ref{fig:SI_Fig_SubHarmonicSim}(b)), they exhibit resonant enhancement much in the same way as conventional soliton dispersive waves \cite{jang_observation_2014}, or in `Kelly'-like sidebands observed in dissipative solitons experiencing a periodic disturbance \cite{nielsen_invited_2018, anderson_dissipative_2023}. We tentatively term these dispersive waves sub-harmonic Kelly sidebands. 

\begin{figure}[H]
	\centering
	\vspace{0pt}
	\includegraphics{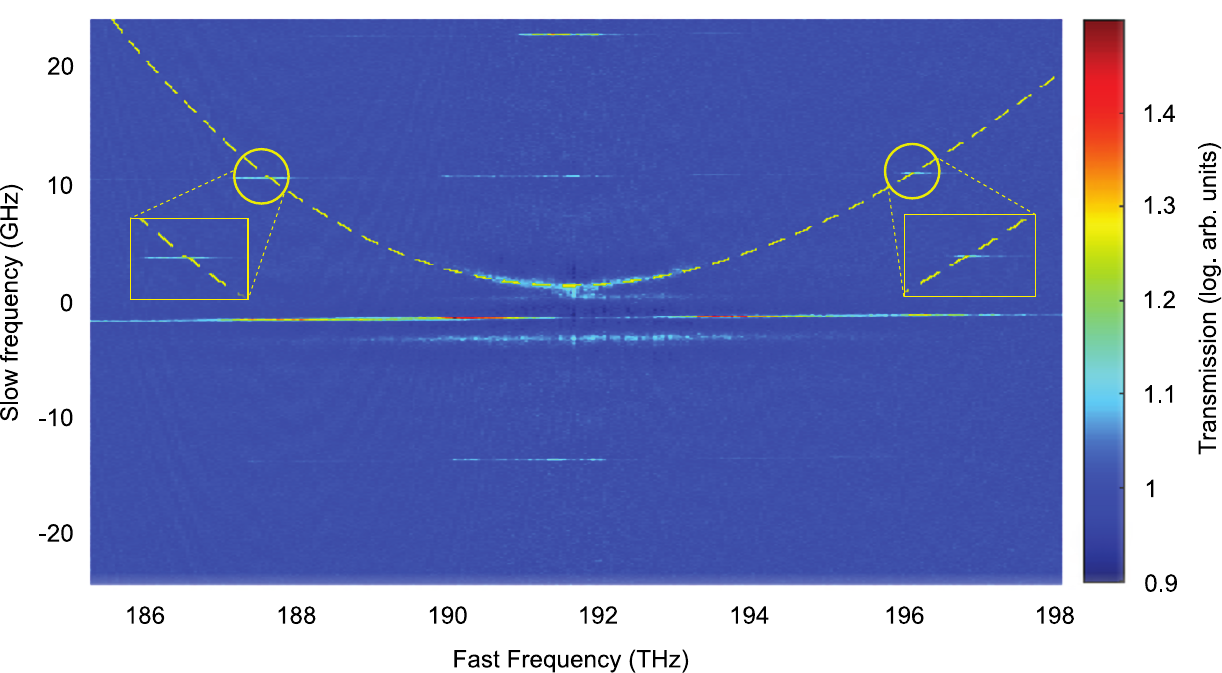}
	\caption{\textbf{Experimental reconstruction measurement of the soliton comb.}
		Each column represents heterodyne detection from one FSR to the next. The dashed line marks the independently measured $D_\mathrm{int}+\delta\omega$ curve. }
	\label{fig:SI_Fig_5}
\end{figure}

Fig. \ref{fig:SI_Fig_5} presents a Kerr comb reconstruction measuring the relative frequencies and linewidths of every comb line based on frequency comb-calibrated swept-laser heterodyne detection \cite{herr_universal_2012}. The latest method for this measurement is described in 
\emph{Tikan et al.} \cite{tikan_emergent_2021}. In this diagram, the subcomb teeth can be seen at $n=-1$, +1, and +2$\times\mathrm{FSR}/4$ relative to the principle soliton comb across the middle. The enhanced subcomb lines are seen to line up exactly with $D_\mathrm{int}+\delta\omega$, where $\delta\omega$ is the laser detuning measured from the VNA, agreeing well with the simulation in Fig. \ref{fig:SI_Fig_SubHarmonicSim}(b) and verifying the Kelly-like sideband theory. The main comb is suppressed around 192~THz due to saturation of the photoreceiver.

\bibliography{references.bib}